\documentclass[lettersize,journal]{IEEEtran}

\usepackage{color,array,amsthm}
\usepackage{graphicx}
\usepackage{amsmath,amsthm, amssymb, latexsym}
\usepackage{upgreek}
\usepackage{subfigure}
\usepackage{orcidlink}
\usepackage{gensymb}
\usepackage{epstopdf}

\begin{document}

\title{Magnetic Field Simulation and Correlated Low-Frequency Noise Subtraction for an in-Orbit Demonstrator of Magnetic Measurements}

\author{Cristian Maria-Moreno\orcidlink{0000-0001-9873-2370}, Ignacio Mateos\orcidlink{0000-0003-3746-9673}, Guillermo Pacheco-Ramos\orcidlink{0000-0001-9655-7820}, Francisco Rivas\orcidlink{0000-0002-6664-4744},\\María-Ángeles Cifredo-Chacón\orcidlink{0000-0002-9620-116X}, Ángel Quirós-Olozábal\orcidlink{0000-0001-6837-9289}, José-María Guerrero-Rodríguez\orcidlink{0000-0002-3765-6498},\\ and Nikolaos Karnesis\orcidlink{0000-0002-2380-3186}

\thanks{Work supported in part by Project PID2022-142281OA-I00 funded by MICIU/AEI/10.13039/501100011033 - FEDER EU, Projects P18-FR-2721 (PAIDI-2020) and FEDER-UCA18-107380 funded by the Dept. of Economy, Knowledge, Business and University, Reg. Governm. of Andalusia, Project IRTP06-UCA funded by the University of Cádiz, and Project AST22\_00001 funded by the European Union - NextGenerationEU, the Dept. of University, Research and Innovation, Reg. Governm. of Andalusia, and the Spanish Ministry of Science and Innovation. The work of I. Mateos was supported by the Ramón y Cajal and Emergia Program. \itshape (Corresp. auth.: I. Mateos)}

\thanks{C. Maria-Moreno, I. Mateos, M.Á. Cifredo-Chacón, Á. Quirós-Olozábal and J.M. Guerrero-Rodríguez are with the School of Engineering, Universidad de Cádiz, Puerto Real, Spain. G. Pacheco-Ramos is with the Dpt. de Ingeniería Aeroespacial y Mecánica de Fluidos, Universidad de Sevilla, Sevilla, Spain. F. Rivas is with the Department of Quantitative Methods, Universidad Loyola Andalucia, Sevilla, Spain. N. Karnesis is with the Department of Physics, Aristotle University of Thessaloniki, Thessaloniki, Greece.}}

\markboth{IEEE Transactions on Instrumentation and Measurement}{}
\maketitle

\begin{abstract}In recent years, nanosatellites have revolutionized the space sector due to their significant economic and time-saving advantages. As a result, they have fostered the testing of advanced instruments intended for larger space science missions. The case of the Magnetic Experiment for the Laser Interferometer Space Antenna (MELISA) is presented in this work. MELISA is a magnetic measurement instrument which aims at demonstrating the in-orbit performance of Anisotropic Magnetoresistance (AMR) sensors featuring dedicated noise reduction techniques at sub-millihertz frequencies. Such low frequency ranges are relevant for future space-borne Gravitational Wave (GW) detectors, where the local magnetic environment of the satellite might yield a significant contribution to the overall noise budget of the observatory. 
The demanding magnetic noise levels required for this bandwidth, down to 0.1 mHz, make measurements arduous. To explore sensing solutions within the H2020 European Commission program with the involvement of the European Space Agency (ESA), the functional performance of MELISA-III will be validated in-orbit. 
During operations, there is the possibility to measure the low-frequency magnetic contribution stemming from orbiting the Earth's magnetic field, which will impede the characterization of the intrinsic performance of the sensor. With the objective of minimizing excess noise during the in-flight operations, the present research aims to simulate the environmental magnetic conditions in LEO in order to identify and subtract undesired contributions to the measurements. The in-orbit long-term magnetic fluctuations are replicated using a tri-axial Helmholtz coil system. A fluxgate magnetometer allows the correlation of the generated field with the payload measurements, leading to the subsequent subtraction. Proving the effect of this approach will facilitate the noise characterization of magnetic sensors in LEO, paving the way for the in-orbit validation of MELISA-III for use in magnetically-demanding missions with long integration times.
\end{abstract}

\begin{IEEEkeywords}CubeSat, electronics, gravitational waves, low frequency, magnetometers, noise, space
\end{IEEEkeywords}

\vfill\null

\section{Introduction}
\IEEEPARstart{T}{he} increasing interest towards the study of GWs has led to the development of dedicated observatories during the last decades. This ongoing scientific effort is spearheaded by the numerous on-ground observations of the advanced GW detector network comprised of the Laser Interferometer Gravitational-Wave Observatory (LIGO), Virgo, and the Kamioka Gravitational Wave Detector (KAGRA) \cite{abbott17}, \cite{abbott19}, \cite{abbott21}, \cite{abbott21b}, \cite{abbott21c}. As a further step in the investigation, extending the observation beyond the Earth's boundaries offers certain advantages. Launching the observatory into space makes it possible to avoid the noisy Earth environment and overcome the length limitations of the interferometer arms, potentially enabling expansion into unexplored lower-frequency regimes. Such research scenarios are pursued through missions like the Laser Interferometer Space Antenna (LISA) \cite{amaro17} by means of the international collaboration between ESA and the National Aeronautics and Space Administration (NASA), as well as the future Chinese GW detectors \cite{gong21}.

The space-borne observatories such as LISA measure GWs by monitoring slight alterations in the relative distance of free-floating test masses inside three identical spacecraft in triangular formation \cite{araujo05}. For the LISA measurement bandwidth, which ranges from $\mathrm{0.1\, mHz}$ to $\mathrm{1\, Hz}$, the test masses are required to present an acceleration noise level below $\mathrm{3\, fms^{-2}Hz^{-1/2}}$ \cite{amaro17}. However, 
environmental magnetic disturbances might lead to an excess of noise, which would violate the test-masses acceleration noise requirements. 

As a result, on-board magnetometers should identify long-term local disturbances as well as the slow drift of the interplanetary magnetic field. These sensors aim to achieve spectral noise densities below $\mathrm{10\, nT \: Hz^{-1/2}}$ along the LISA bandwidth \cite{mateos15b}, \cite{mateos15}.

For that reason, it is crucial to demonstrate the technology featured in high-budget missions \cite{sauser2006}.

In recent times, CubeSats have proved to be an ideal platform for technology demonstrators \cite{{hubert2011}}, leading to affordable improvements in the Technology Readiness Level (TRL) of key instruments through validation in an orbit environment.

There lies the reason for the development of MELISA, a CubeSat payload featuring magnetic-shielded AMR sensors with dedicated electronic noise reduction techniques \cite{mateos15}, \cite{mateos18}, \cite{mateosthesis}. A version of this magnetic measurement system, MELISA-III, is onboard the $\Upsigma$yndeo-2 nanosatellite as part of the Horizon 2020 programme \cite{mateos2023}. An ESA Vega launcher has successfully injected the satellite into a Sun Synchronous Orbit (SSO) in October 9th 2023 from the ESA's Spaceport in French Guiana. 
In-orbit operations are expected to begin in 2024. Therefore, the work presented here uses this payload version for the experimental assessments.

Additionally, a variation of this measurement system, MELISA-II, is included in UCAnFly, a CubeSat developed under the Fly Your Satellite! (FYS) programme of ESA and whose launch is expected by late 2025.

In the present work, we simulate and generate the LEO magnetic conditions in order to identify and subtract the correlated undesired magnetic effects in the measurements, with the objective of assessing just the intrinsic noise of the payload. In the context of low-frequency field simulations, unwanted contributions that are not present at higher regimes may appear, such as magnetic fluctuations due to the material behaviour.

For instance, the thermal variations experienced in the environment impact the magnetic shield properties and the electronics, resulting in the need to correlate the measurements with the payload's temperature sensor as well. The effectiveness of the proposed simulator scheme and the noise subtraction procedure is evaluated for such low frequencies not addressed for LEO magnetic field simulators. The validation of this procedure will be vital for the future on-ground postprocessing, in which the in-orbit outputs from the external sensors of the $\Upsigma$yndeo-2 nanosatellite will be considered for the correlation. The sub-millihertz evaluation of the generated field will also help understanding the behaviour and limitations of the long-term simulations of LEO magnetic fields.

The study of magnetic field simulators has been done in the past \cite{uscategui23,deloiola18,dasilva19,piergentili11,takahiro14}.

However, those works do not encompass the assessment of the low-frequency behaviour of the generated field, lacking noise characterization of the generated field down to 0.1 mHz. They often focus on procedures for testing the Attitude Determination and Control System (ADCS) of nanosatellites. Besides, the subtraction of the magnetic contributions measured by different sensors requires additional considerations, such as minimizing the quantification of the generated field and synchronizing the readouts of the different sensors. Additionally, in this research, the in-orbit attitude of the satellite is already considered in the simulated field by means of coordinate transformations to fit the $\Upsigma$yndeo-2 body-centered frame.
 
The subtraction of magnetic noise in the context of interferometry-based missions has been studied before~\cite{coughlin16,thrane14,coughlin18}. For those cases, the use of Wiener filtering \cite{{allen99}} allows the removal of environmental contributions using reference signals from external sensors. Here, conversely, the conditions of the magnetic sensors within the $\Upsigma$yndeo-2 nanosatellite do not allow a simple correlation because the MELISA-III's sensors are shielded, in contrast to the rest of the spacecraft's magnetometers. 
This design detail, together with the aforementioned differences in the generation of the LEO magnetic field, entails the development of an updated approach suitable for this kind of in-orbit scenarios.

This work is organized as follows: Section II describes the magnetic measurement system used throughout this research, MELISA-III; Section III elaborates on the methodology followed for the simulation of the orbit environment and the generation of the LEO magnetic field; Section IV shows the procedure that allows the subtraction of correlated noise; in Section V, the impact of using a different magnetic shield attenuation on the noise subtraction is analyzed; Section VI includes the results of the field simulation, generation, and noise subtraction from the data provided by the sensors; finally, Section VII presents the conclusions of this work.

\section{The MELISA-III Magnetic Measurement Demonstrator}

\begin{figure}[!t]
\centering
\subfigure[]{\includegraphics[width=0.5\linewidth]{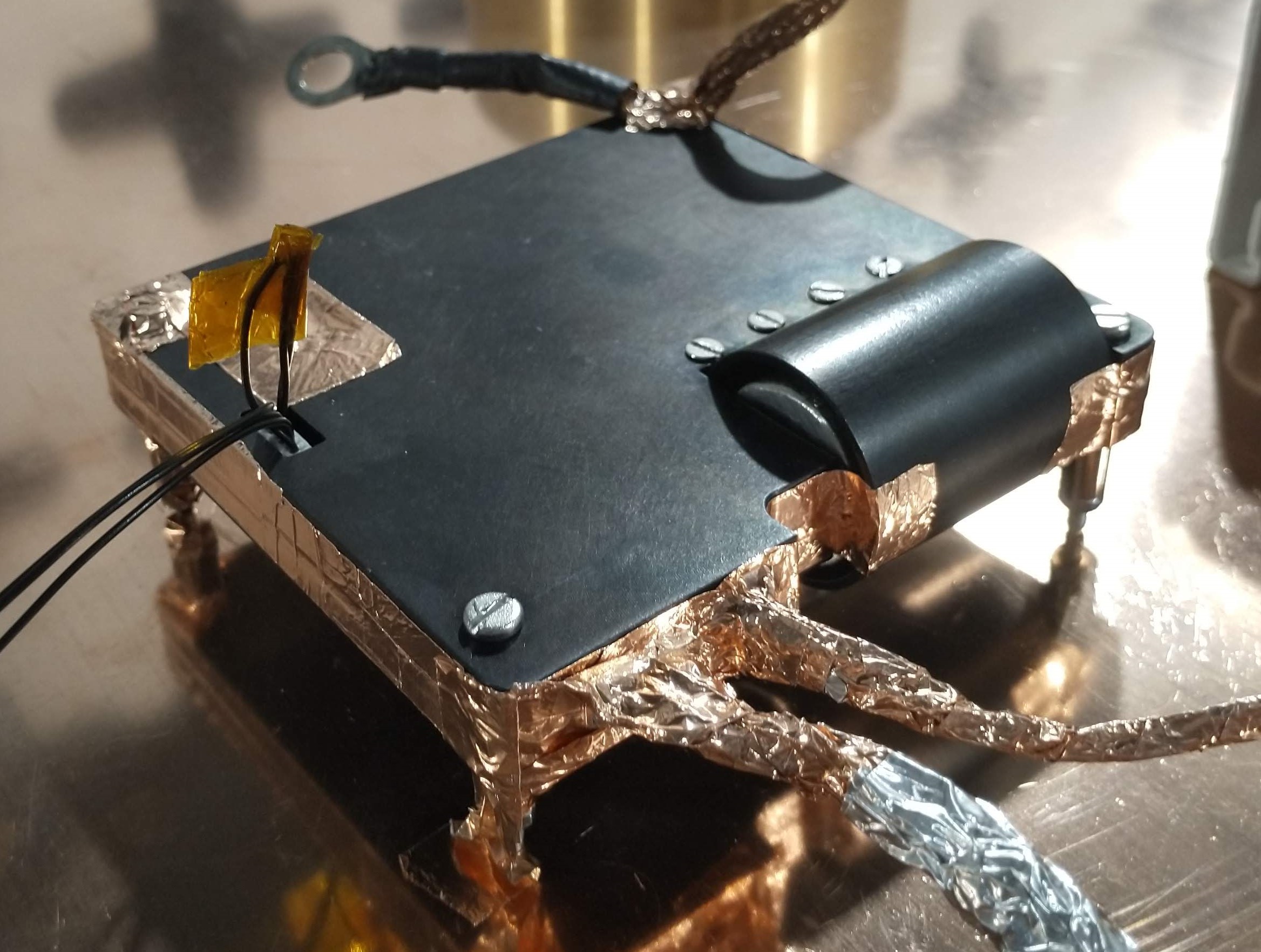}}
\hspace{0.5 cm}
\subfigure[]{\includegraphics[width=0.41\linewidth]{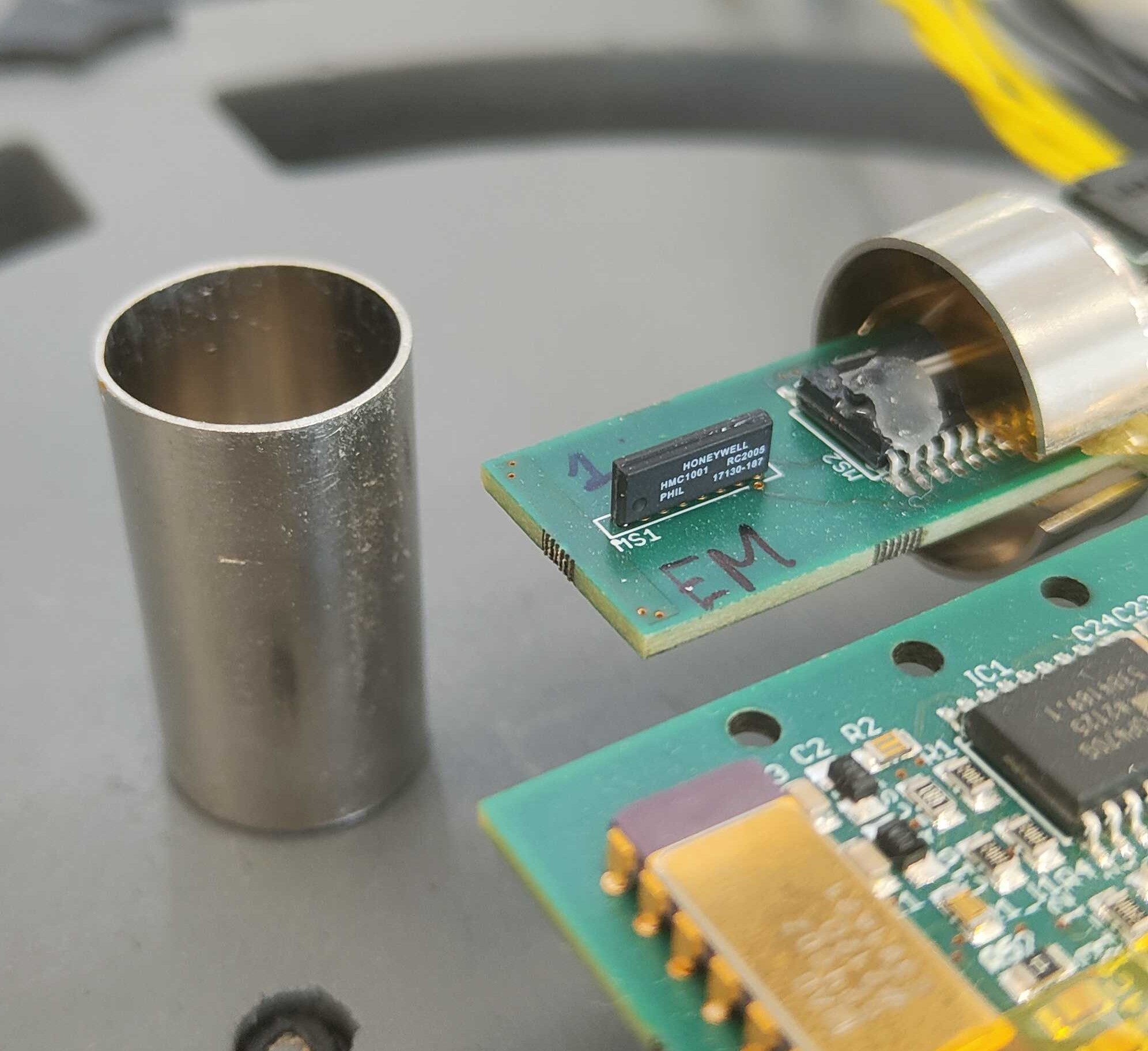}}
\caption{(a) FM of MELISA-III during the qualification campaign and (b) the one-layer magnetic shield used for the present experiment.}
\label{fig:melisa}
\end{figure}

The payload, MELISA-III, consists of a PC104 form factor PCB featuring a uniaxial (HMC1001) and a biaxial (HMC1002) AMR sensors placed inside a three-layer cylindrical shield of mu-metal (nickel-iron alloy) in order to attenuate external magnetic contributions \cite{wadey56}, having its physical representation displayed in Fig. \ref{fig:melisa}.  

\begin{table}[!t]
\centering
\caption{Payload's cylindrical mu-metal shield description.}
\vspace{-5pt}
\begin{tabular}{l  c c}
\hline \hline
{\bf Parameter} & {\bf Value}\\ 
	\hline 
	Alloy & Co-NETIC AA \cite{magneticshield} \\ 
	Composition (weight) & 80\% Ni; 4.9\% Mo; Fe balance; 0.5\% Mn; 0.3\% Si  \\
	Permeability & 15000 - 25000 (for the sub-millihertz range) \\
    Thickness & 0.64 mm \\
    Diameter & 19.79 mm \\
    Length & 35.45 mm \\
	\hline
	\hline
\end{tabular}
\label{tab:mumetal}
\end{table}

\begin{figure}[!t]
\centering
\includegraphics[width=1\linewidth]{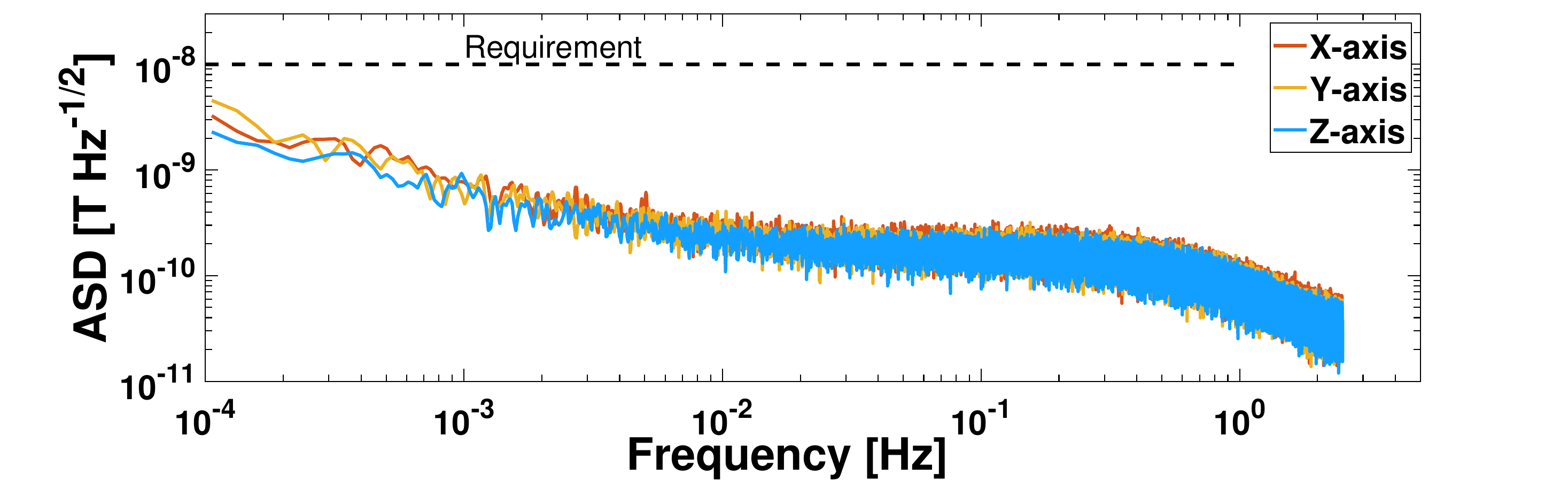}
\caption{ASD for MELISA-III's measurements with the three-layer magnetic shield that will be featured during the in-orbit operations.}
\label{fig:melisa_ASD}
\end{figure}

Its analog and digital circuits allow magnetic field and temperature measurement, noise reduction and external communication. Under laboratory conditions, the system has proved to be compliant with the noise level requirements in the LISA bandwidth, shown in Fig. \ref{fig:melisa_ASD}. As planned, the in-orbit demonstration of this technology would increase the maturity of the system from TRL 4 to TRL 6 \cite{amaro17,mankins1995}. It is worth mentioning that, since the three-layer magnetic shield was already put into orbit with the Flight Model (FM) of MELISA-III, the one available for the presented research is the one-layer model with lower attenuation (described in Table \ref{tab:mumetal}), manufactured for the alternative payload version, MELISA-II. This shield consists of a cylindrical layer with a cap placed at each end. One cap is welded to the body of the cylinder, while the other is removable and presents a slot to fit into the PCB. The payload’s X-axis is aligned with the normal direction of the caps, thereby reducing the effectiveness of the shield in attenuating external fields along this axis.

Nonetheless, when it comes to performing the payload validation in a LEO environment, a magnetic shield might not be sufficient. The low-frequency magnetic contributions to the payload's measurements originated from orbiting the Earth's magnetic field may increase the overall noise levels as described below for the amplitude spectral density:

\begin{equation}
 S_{\rm meas}^{1/2} = \sqrt{ S_{\rm FEE} + S_{\rm LEO} + S_{\rm SAT} + (\upalpha_{\rm FEE}^2 + \upalpha_{\rm shield}^2) S_{\rm T}}
  \label{for:nose_density}
\end{equation}

\noindent where $S_{\rm FEE}$ is the noise power spectral density related to the front-end electronics; $S_{\rm LEO}$ is the Earth's magnetic field power spectrum originated from orbiting in LEO; $S_{\rm SAT}$ is the magnetic field spectrum caused by the local magnetic sources of the satellite; $\upalpha_{\rm FEE}$ and $\upalpha_{\rm shield}$ are the thermal coefficients of the front-end electronics and the magnetic shield; and $S_{\rm T}$ is the power spectral density due to thermal fluctuations. In the measurement bandwidth, the effects of $S_{\rm FEE}$ and $S_{\rm SAT}$ do not entail exceeding the noise requirements, whereas $S_{\rm LEO}$ and $S_{\rm T}$ do, requiring subtraction to remove excess fluctuations. Since they are considered independent errors in the measurement, the total amplitude spectral density can be defined as the square root of the individual power spectral densities \cite{pallas1999}.

\section{Simulation and Generation of the LEO Magnetic Field}

\subsection{LEO magnetic environment simulation}

\begin{figure}[!t]
\centering
\includegraphics[width=0.5\linewidth]{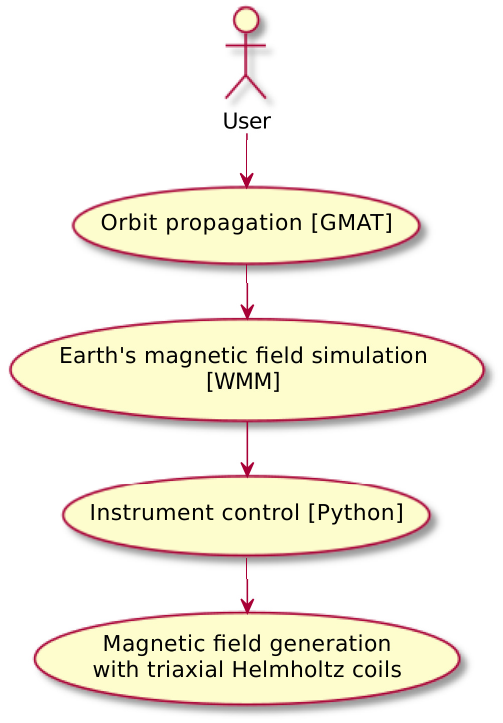}
\caption{Sequence for the simulation and generation of the orbit magnetic field.}
\label{fig:sequence}
\end{figure}

The methodology followed to replicate the space conditions in which the payload's in-flight operations will be performed is depicted in Fig. \ref{fig:sequence}, starting with the orbit definition. Since the $\Upsigma$yndeo-2 satellite has been launched into a SSO, the expected orbital parameters that were provided by the launcher at the start of this work are detailed in Table \ref{tab:orbit_elements}, along with the actual values of the launch. The mismatch between the simulated and the launch values is acceptable since different scenarios have been tested (6942.1 km $\pm$ 75 km) and each one resulted in a noise contribution at the same frequency. Regarding the attitude, the Z+ direction of the satellite is nadir pointing, while X+ is aligned with the anti-velocity vector and Y+ is parallel to the orbit plane normal.

\begin{table}[!t]
\centering
\caption{Description of SSO for the $\Upsigma$yndeo-2 Satellite.}
\vspace{-5pt}
\begin{tabular}{l  c c}
\hline \hline
{\bf Parameter} & {\bf Simulated values} & {\bf Launch values} \\ 
	\hline 
	Semi-major axis & 6942.100 km & 6940.8 km\\ 
	Eccentricity & 0.00120 & 0.00028\\
	Inclination & 97.66 deg & 97.7 deg\\ 
	\hline
	\hline
\end{tabular}
\label{tab:orbit_elements}
\end{table}

A simulated trajectory is then propagated from the previous initial conditions using GMAT, the open-source software provided by NASA for space mission design, optimization, and navigation \cite{hughes16}. The atmospheric model used is the Mass Spectrometer Incoherent Scatter (MSIS)E90 \cite{hedin87}, while the gravity model is the Joint Earth Gravity Model (JGM)-2 \cite{nerem94}. With the calculated position of the spacecraft over time, a standard model for the geomagnetic field (WMM) provides the magnitude and direction of the Earth's magnetic field contribution at each point along the trajectory \cite{mchulliat}.

\begin{figure}[!t]
\centering
\includegraphics[width=1\linewidth]{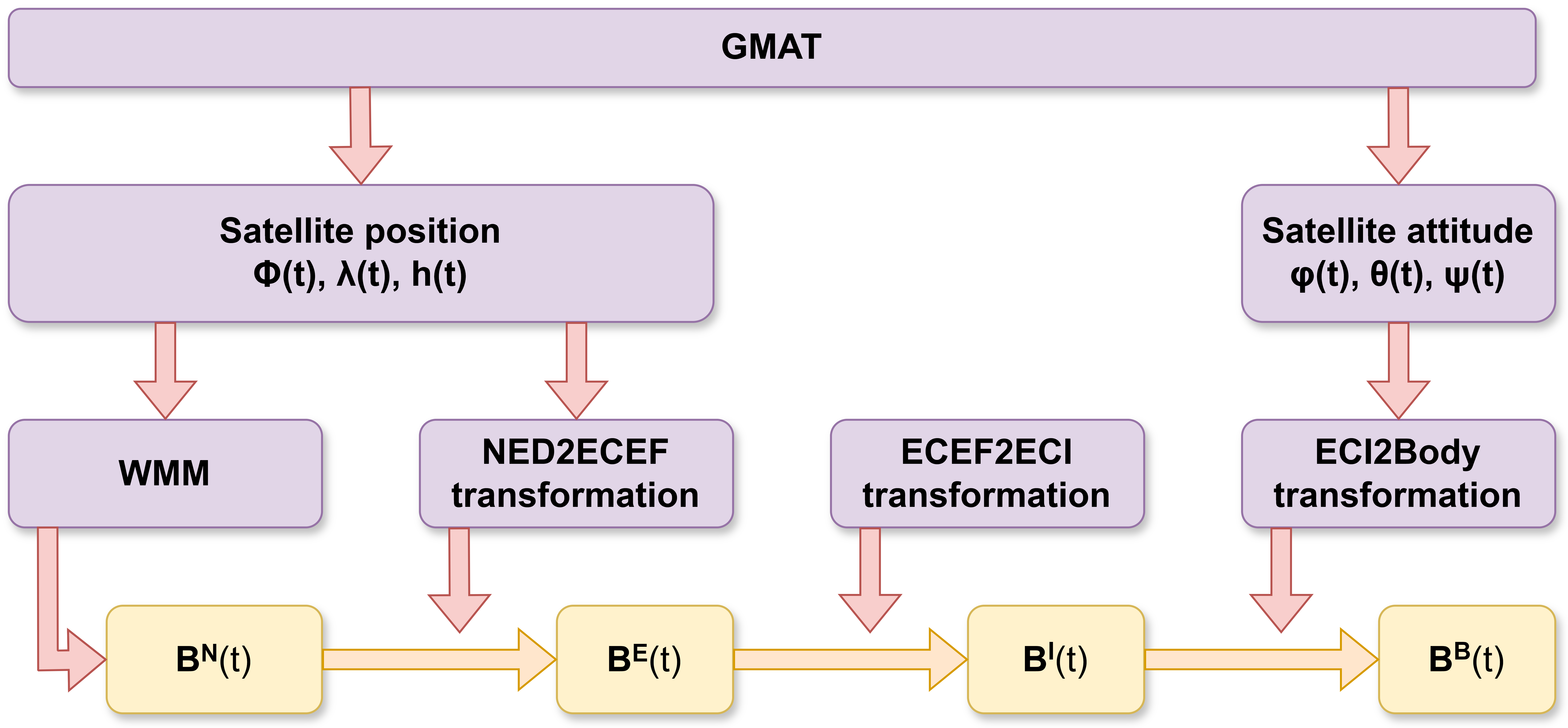}
\caption{Transformations between coordinate systems from NED reference frame to satellite body frame.}
\label{fig:transformations}
\end{figure}

It should be noted that different transformations between coordinate systems have to be carried out as illustrated in Fig. \ref{fig:transformations}, where the superscript of the magnetic field indicates the reference frame in which it is expressed. Given that the WMM provides the geomagnetic vector expressed in the North East Down (NED) reference frame, and that MELISA’s AMRs are aligned with the satellite body frame (B), the conversion into this last frame of reference is needed. Based on the results obtained from GMAT for the evolution of the geodetic coordinates ($\Upphi$, $\uplambda$ and $\mathrm{h}$) establishing the satellite position and the Euler angles ($\upphi$, $\uptheta$ and $\uppsi$) defining its attitude, successive transformations are performed.  The magnetic field vector, initially expressed in the NED frame $\mathrm{B^N}$, is then transformed to the Earth-Centered, Earth-Fixed frame (ECEF), the Earth-Centered Inertial frame (ECI) and finally to the Body frame (B).

\subsection{Orbit magnetic field simulator scheme}\label{sec:magnetic_field_generation}

\begin{figure}[!t]
\centering
\subfigure[]{\includegraphics[width=1\linewidth]{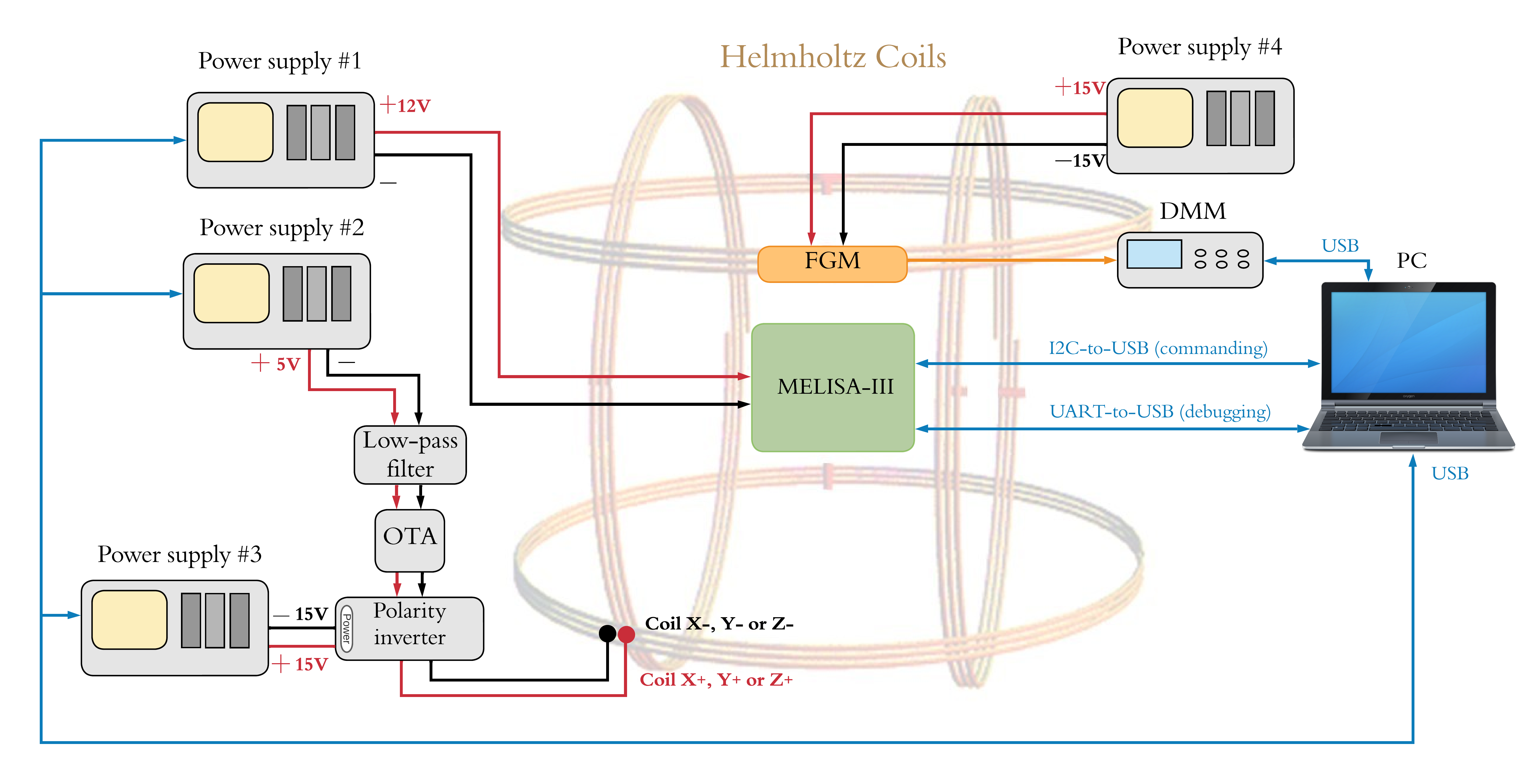}}
\subfigure[]{\includegraphics[width=0.9\linewidth]{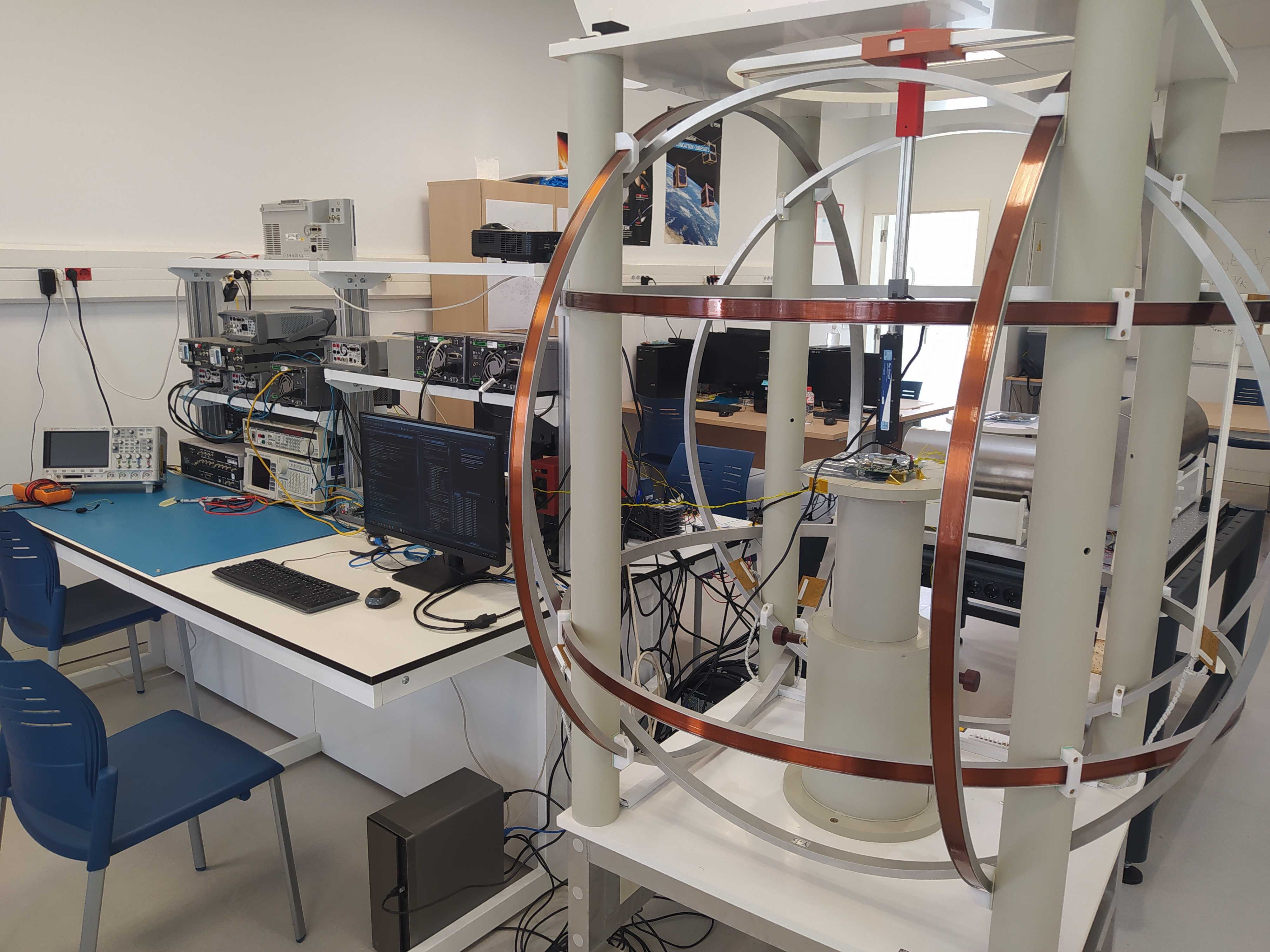}}
 \caption{(a) Block diagram of the setup and (b) physical representation of the setup in the laboratory, with the payload and the FGM located inside the Helmholtz coils.}
 \label{fig:setup}
\end{figure}

To generate the previously simulated magnetic environment, MELISA-III is placed inside a tri-axial Helmholtz coil system \cite{serviciencia} as displayed in Fig. \ref{fig:setup}. The coil's field-to-current ratio, which has been calibrated with an optically pumped atomic magnetometer \cite{quspin}, is defined as $200 \ \upmu \mathrm{T \: A^{-1}}$ per axis. A programmed power supply provides the required current to the coils, while a polarity inverter enables the generation of negative values of magnetic field by reversing the current's direction. The power supply output is controlled via GPIB using a Python script, which generates the in-orbit magnetic values obtained from the WMM earlier. However, the quantized signal produced by the control of the equipment induces high-frequency noise into the generated field. Thus, incorporating a passive low-pass filter, in combination with an Operational Transconductance Amplifier (OTA), allows the smoothing of magnetic amplitude variations and reduce the quantization noise. The OTA features an OPA544 power operational amplifier \cite{texas}, performing the V-I conversion with floating loads. The cut-off frequency of the first-order filter is set to $\mathrm{40 \, mHz}$, lower than the frequency at which this noise appears (100 mHz).

For the correlation of MELISA-III's measurements, a Fluxgate Magnetometer (FGM) \cite{bartington} is also located within the homogeneous magnetic field volume of the coil. Conversely to the payload, the FGM is not shielded, so the same configuration as for the external magnetometer of the $\Upsigma$yndeo-2 is replicated. Inside the mu-metal shield that surrounds the AMR sensors in MELISA, a thermistor is used to monitor low-frequency temperature fluctuations that can affect the payload's magnetic readouts.

Ensuring synchronization between the magnetic measurements of MELISA-III and the FGM is crucial for accurate noise correlation. Therefore, FGM magnetic data are triggered from an external source and gathered by a 7.5-digit Digital Multimeter (DMM) \cite{keysight} at a sampling frequency of ${5 \, {\rm Hz}}$. The temperature sensor, on the other hand, is acquired through MELISA-III's Analog-to-Digital Converter (ADC) at ${5 \, {\rm Hz}}$ as well, allowing for synchronization with the magnetic data. After simulating the orbital cycles, the magnetic measurements of the FGM and the NTC temperature information enable the correlation of the induced noise contributions to MELISA-III's scientific data.

\section{Noise Subtraction Procedure}
\label{sec:analysis}

Once the magnetic field has been collected, the non-shielded FGM's measurements are fitted using least squares in the frequency domain to the shielded payload's data in the bandwidth impacted by the orbit simulation, around $\mathrm{0.1 \, mHz}$. The objective is to reduce the FGM's amplitude in order to match the attenuation effect of the AMR sensors' magnetic shield. Considering a linear relationship, we use a single-parameter model, $\mathrm{B_M(t) = c_A \times B_F(t)}$, where $\mathrm{B_M}$ and $\mathrm{B_F}$ are the measurements of MELISA-III and the FGM in Tesla, and $\mathrm{c_A}$ is the attenuation to be estimated. Afterwards, the fitted curve is subtracted from the payload's measurements.

The analysis procedure follows the Iterative Reweighted Least Squares algorithm introduced in \cite{vitale14} and used extensively during the LISA Pathfinder mission operations \cite{armano18}. The software is  part of the LISA Technology Package Data Analysis (LTPDA) toolbox \cite{ltpda}. We first define a linear model $h(\vec{\theta}; t)$, which depends on a parameter set $\vec{\theta}$, and then form the residuals time series $r(t) = d(t) - h(\vec{\theta}; t)$. In our particular case here, the situation greatly simplifies since our model is a single parameter that describes the linear relation between the two time-series measured by the two magnetometers. Then, we iteratively compute $\chi^2_n$ as
\begin{equation}
    \chi^2_n= N_s\sum_{j\in Q}\frac{ \overline{\left| \tilde{r}_j (\vec{\theta}_n)\right|} }{ \overline{\left| \tilde{r}_j (\vec{\theta}_{n-1})\right|} },
    \label{eq:flscov}
\end{equation}
using the residuals at the previous iteration $n-1$, until we reach convergence (no change on the parameter values up to a given pre-defined threshold). Here, the $(\tilde{\,\,})$ represents the transformation to the frequency domain, $j$ represents the given frequency bin, and $N_s$ the number of data stretches used to perform the averaging for the computation of the power spectrum of $r$. In our particular case, since the model is quite simple (see section~\ref{sec:results}), the procedure based on eq.~(\ref{eq:flscov}) becomes a least square fit in frequency domain, where the parameters errors are computed by matrix inversion (we refer the reader to~\cite{vitale14} for more details of this methodology).

A similar procedure to the one described above is used for the thermal subtraction. Temperature fluctuations in the laboratory may increase the noise below 1 mHz. Therefore, the temperature data provided by the NTC thermistor included in MELISA-III are also fitted to the payload's measurements for a second noise correlation. Consequently, the combination of both subtraction procedures will allow us to reduce the environmental noise at lower frequencies and unveil the intrinsic noise of the sensors. The correlation between two different signals will be quantified using the Pearson's Product-Moment Correlation Coefficient (PPMCC) \cite{puth14}.

\section{Simulating a Different Shielding Attenuation}

\begin{figure}[!t]
\centering
 \includegraphics[width=0.9\linewidth]{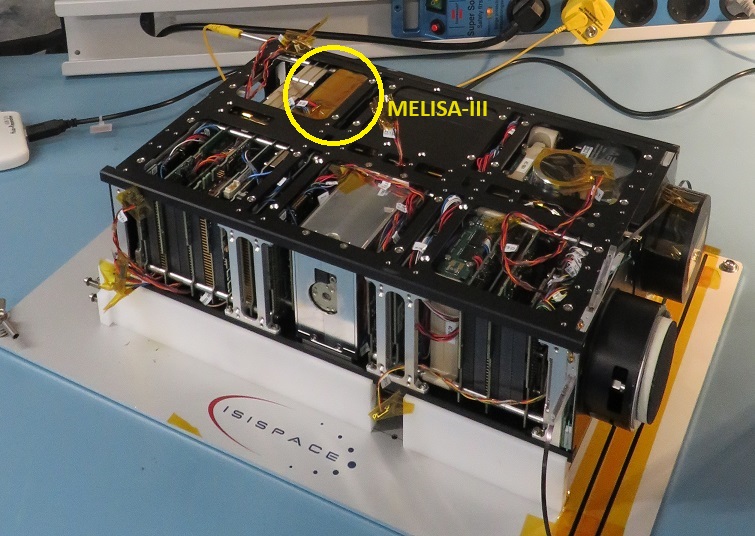}
 \caption{Mu-metal box that contains MELISA-III during the integration in the $\Upsigma$yndeo-2 platform, performed by the system integrator ISISPACE.}
 \label{fig:integration}
\end{figure}

As outlined earlier, the subtraction procedure is accomplished using a one-layer magnetic shield for the AMR-sensors. However, for the actual in-orbit operations of MELISA-III, the external magnetic field will be further attenuated due to the featured three-layer shield and an external mu-metal box that the CubeSat integrator has used for surrounding the payload, as depicted in Fig. \ref{fig:integration}. Then, the greater attenuation provided by the whole set of shields incorporated in the satellite will result in an enhanced noise reduction. To estimate the benefits in the subtraction process, the actual magnetic field that MELISA-III experiences in orbit will be replicated in the laboratory by reducing the amplitude of the field generated with the Helmholtz coils.

Obtaining the amplitude of the attenuation within the satellite is not straightforward, given the absence of the magnetic characterization of the mu-metal box included in the spacecraft. Even so, the measurements of MELISA-III during the CubeSat integration may offer an insight into the actual value of field reduction as long as the local magnetic field of the $\Upsigma$yndeo-2 is known. 

The dominant local magnetic source of the satellite originates from a different experiment, a radially magnetized permanent magnet ring integrated near the payload. Its magnetic moment follows the next expression:

\begin{equation}
m=\frac{B_{\mathrm{r}}}{\upmu_{\mathrm{0}}}dV=\frac{B_{\mathrm{r}}}{\upmu_{\mathrm{0}}}(r-b)ldp
\end{equation}

\noindent where $B_{\mathrm{r}}$ is the remanence of the material; $\mu_{\mathrm{0}}$ is the magnetic permeability constant; $r$, $b$ and $l$ are the outer radius, inner radius, and thickness of the ring. Therefore, from the readouts of MELISA-III and the predicted field generated by the magnet in the location of the payload, the real attenuation after integration can be estimated.

\section{Results}

\subsection{In-orbit Earth's magnetic field contribution}

\begin{figure}[!t]
\centering
\includegraphics[width=1\linewidth]{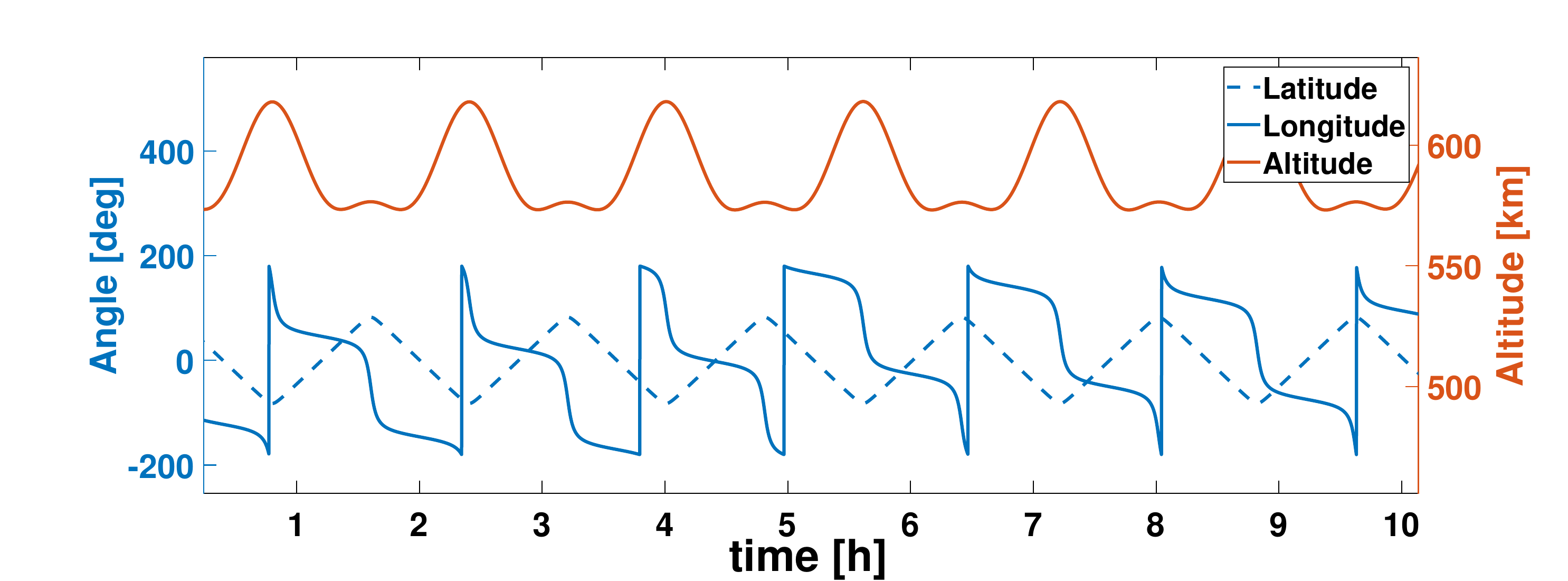}
\caption{Latitude (dotted blue, in deg), longitude (solid blue, in deg), and altitude (red, in km) of the spacecraft for 10 hours during the orbit simulation.}
\label{fig:gmat_3}
\end{figure}

\begin{figure}[!t]
\centering
\includegraphics[width=1\linewidth]{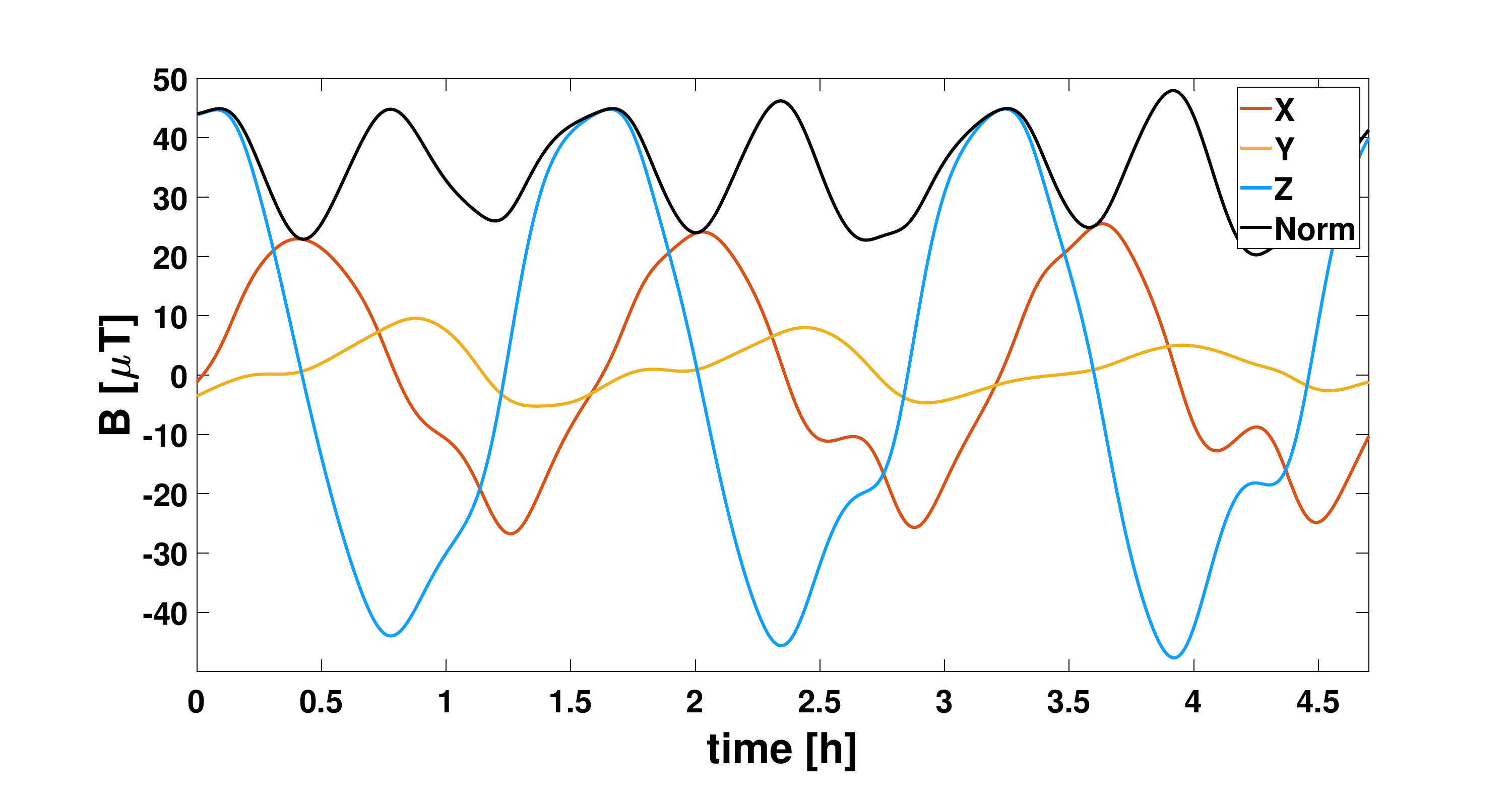}
 \caption{Orbit simulation Earth's magnetic field components expressed in body axes: X-axis (red), Y-axis (orange) and Z-axis (blue), along with the norm (black).}
 \label{fig:earth_field}
\end{figure}

During the simulation of the orbit, the latitude, longitude, and altitude values of the satellite are collected as shown in Fig. \ref{fig:gmat_3}. Then, the environmental field that surrounds the satellite for 5 hours of simulation is displayed in Fig. \ref{fig:earth_field}.

\begin{figure}[!t]
\centering
\includegraphics[width=1\linewidth]{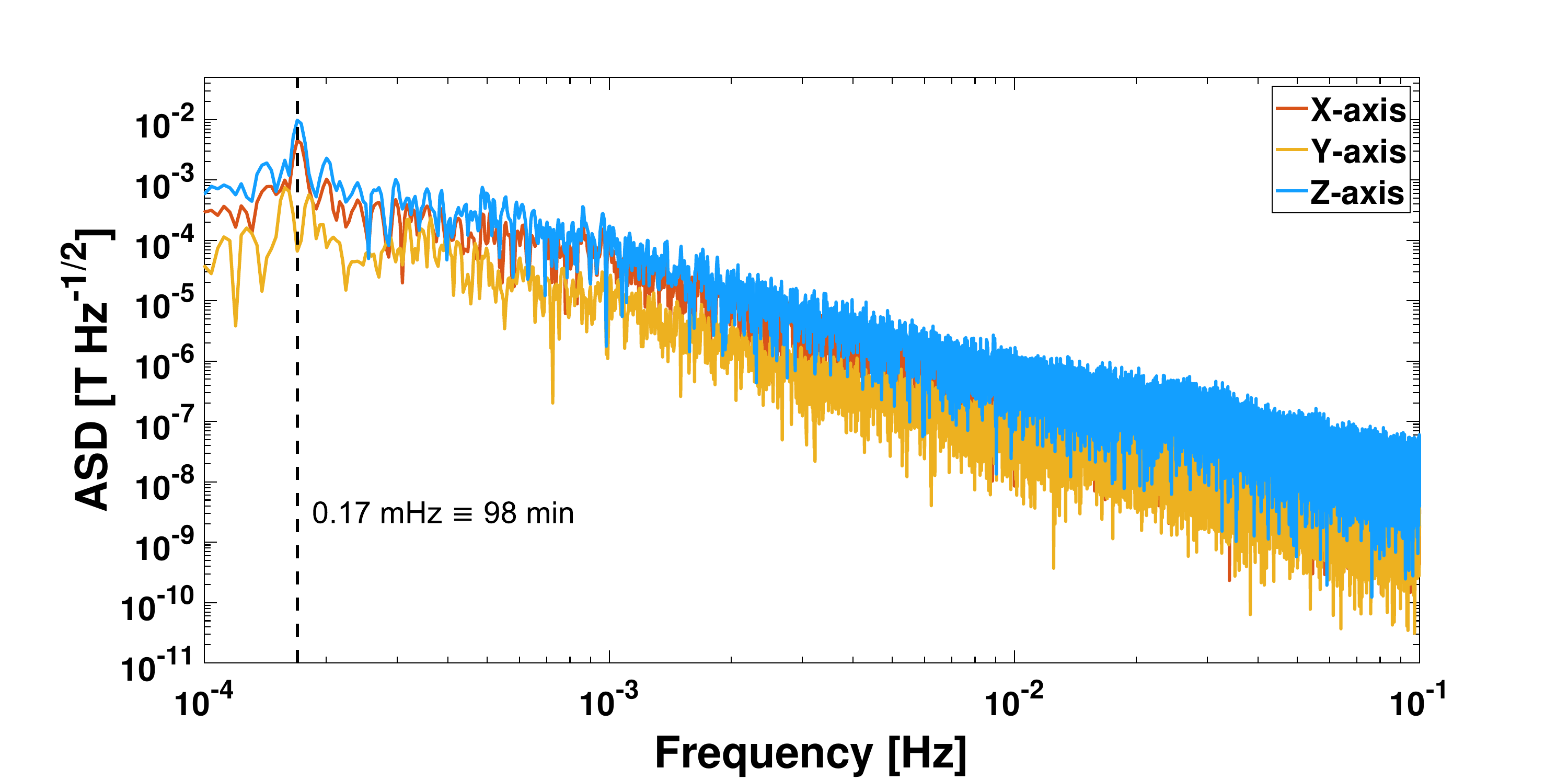}
 \caption{ASD of the simulated Earth's magnetic field contributions throughout the orbit in the X-axis (red), Y-axis (orange) and Z-axis (blue) of the spacecraft. The peak at 0.17 mHz corresponds to the fundamental frequency of the magnetic field noise contributions generated by orbiting the Earth.}
 \label{fig:asd_earth}
\end{figure}

In order to reduce the errors in the estimate of the spectral densities within the targeted bandwidth, it is necessary to simulate at least three days of orbital cycles. Afterwards, the environmental noise levels associated with the Earth's magnetic field with respect to each axis of the spacecraft are represented through the Amplitude Spectral Density (ASD) plot \cite{ltpda} displayed in Fig. \ref{fig:asd_earth}. The pronounced peak at $\mathrm{0.17 \, mHz}$ corresponds to the fundamental frequency of the magnetic fluctuations throughout the orbit, whose period is approximately 98 min.

\subsection{Magnetic field generation}

\begin{figure}[!t]
\centering
 \includegraphics[width=1\linewidth]{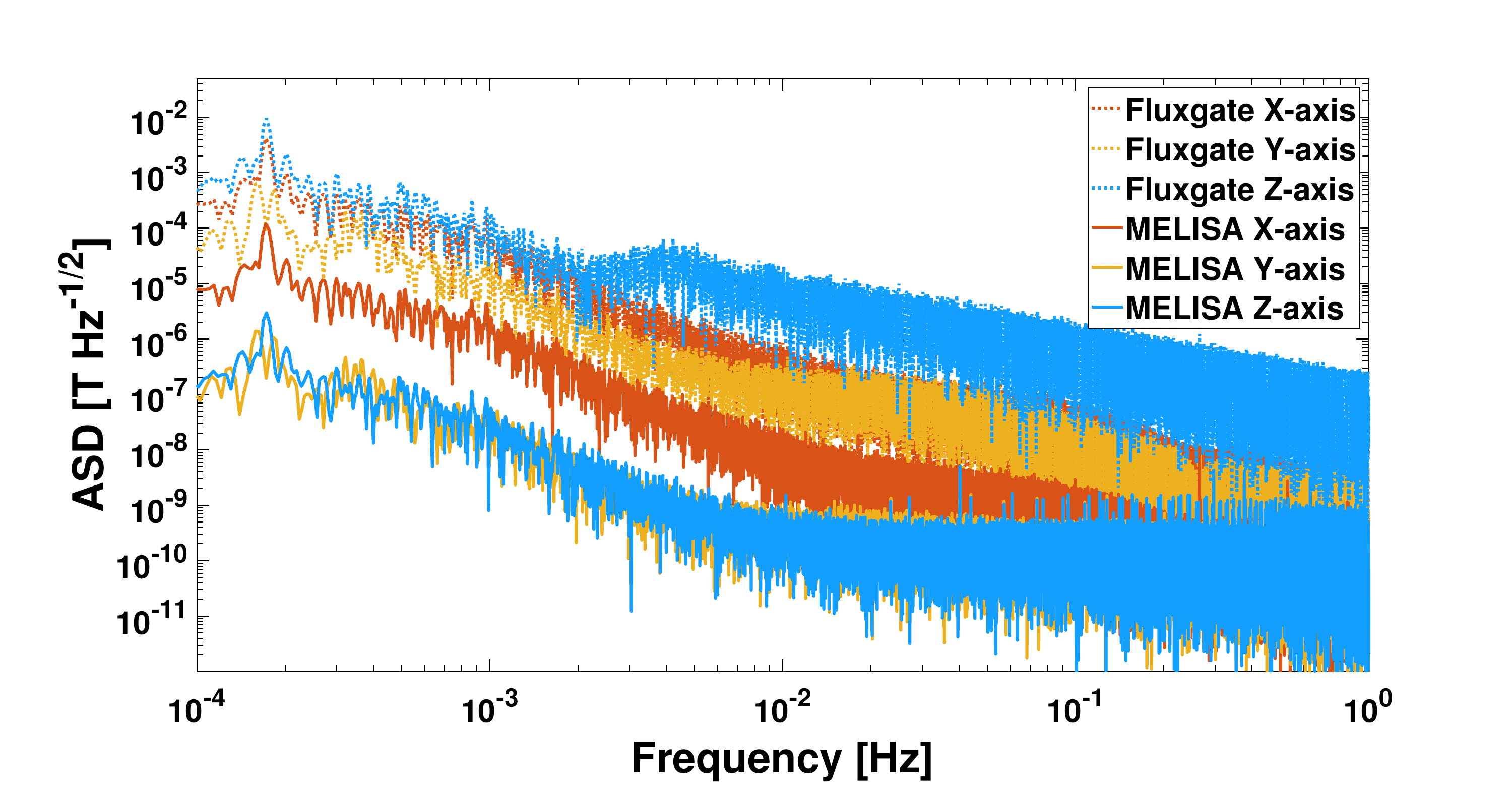}
 \caption{ASD of the FGM and MELISA-III measurements during the generation of the orbit magnetic environment for the X-axis (red), Y-axis (orange) and Z-axis (blue).}
 \label{fig:helmholtz_low}
\end{figure}

Previous data have been converted to the current applied to the Helmholtz Coils to generate the expected magnetic environment in orbit. The measurements of the payload and the FGM are displayed in the frequency domain in Fig. \ref{fig:helmholtz_low} and in the time domain in Fig. \ref{fig:with_without_temp}. As mentioned before, the ambient field collected by MELISA-III is attenuated with respect to the FGM due to the magnetic shield that surrounds the payload's AMR sensors. That reduction is different for each axis due to the variations in the shielding attenuation based on the direction. The Y- and Z-axes exhibit a negative time-domain correlation because the magnetometers' positive senses are opposite. Since peaks at the frequency of the orbital period are present in the measurements of MELISA-III and the FGM, a correlation between both sensors will allow us to subtract this induced noise in the payload's scientific data. Further details are provided in the next section.

To validate the magnetic generation process, we compare the actual field measured by the non-shielded sensor with the results of the previous simulations in Table \ref{tab:error}. The relative error in all the directions is less than $5\,\%$, making the estimated accuracy acceptable for subsequent noise subtraction.

\begin{table}[!t]
\centering
\caption{Comparison between the Predicted Peak-to-peak Amplitude for the Simulated In-orbit Magnetic Field and the Generated Magnetic Field Measured by the FGM.}
\vspace{-5pt}
\label{table}
%\tablefont
\begin{tabular}{cccc}
\hline \hline
\textbf{Axis} & \textbf{\begin{tabular}[c]{@{}c@{}}Predicted in-orbit\\ magnetic field,\\ max peak-to-peak\\ {[}$\upmu \mathrm{T}${]}\end{tabular}} & \textbf{\begin{tabular}[c]{@{}c@{}}Generated\\ magnetic field,\\ max peak-to-peak\\ {[}$\upmu \mathrm{T}${]}\end{tabular}} & \textbf{\begin{tabular}[c]{@{}c@{}}Error\\ {[}\%{]}\end{tabular}} \\
\hline
X & 61.59 & 58.60 & 4.85 \\
Y & 22.70 & 22.07 & 2.78 \\
Z & 95.92 & 93.82 & 2.19 \\
\hline \hline
\end{tabular}
\label{tab:error}
\end{table}

\subsection{Noise subtraction outcomes}\label{sec:results}

\subsubsection{Magnetic field contribution}

\begin{figure}[!t]
\centering
 \subfigure[]{\includegraphics[width=1\linewidth]{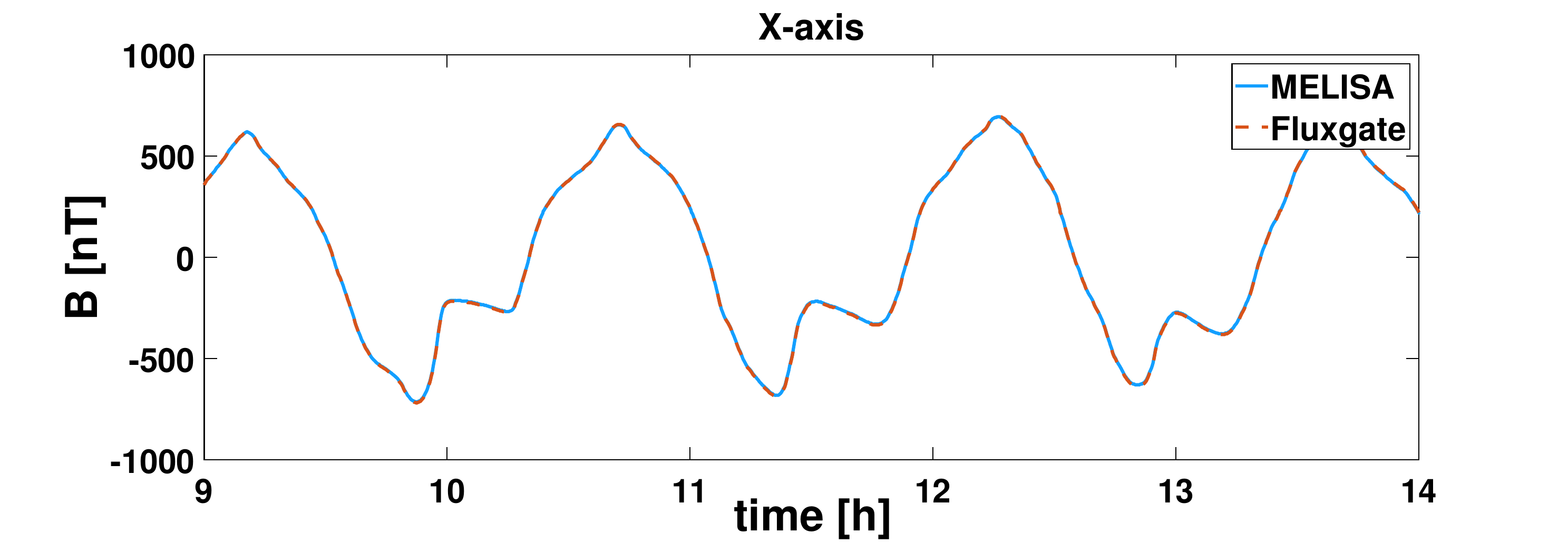}}
 \subfigure[]{\includegraphics[width=1\linewidth]{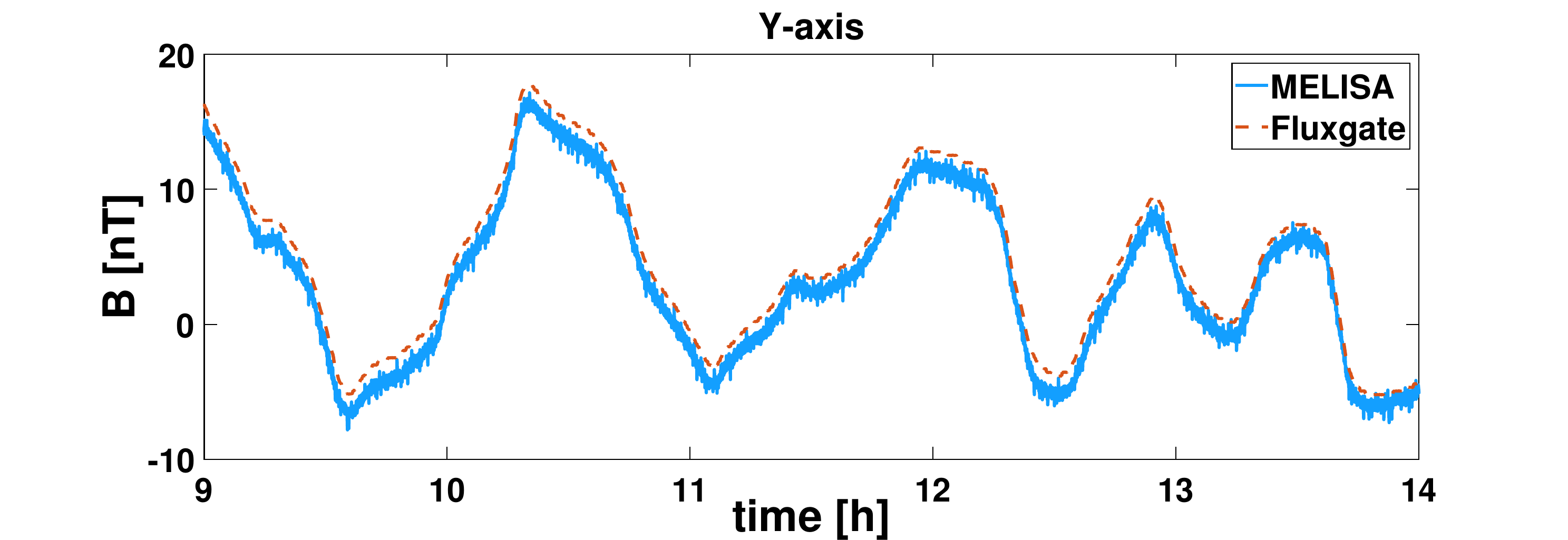}}
 \subfigure[]{\includegraphics[width=1\linewidth]{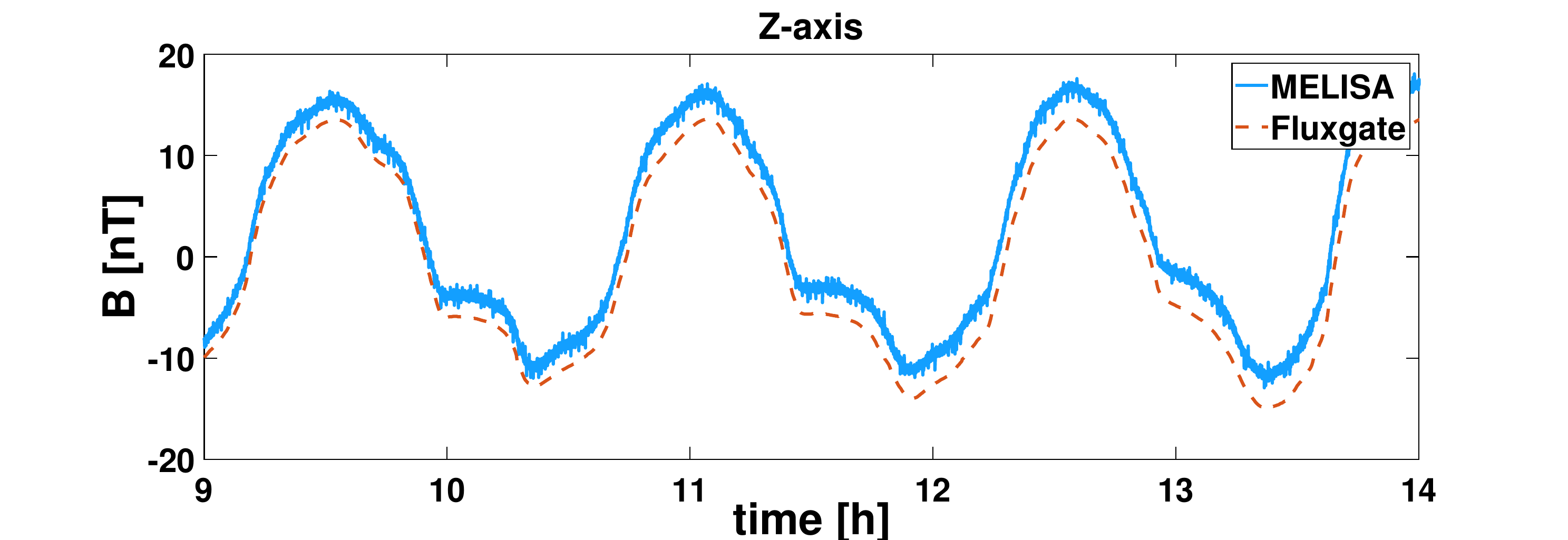}}
 \caption{Fit applied to the FGM's measurements to match MELISA-III's readouts after detrending in the (a) X-axis, (b) Y-axis, and (c) Z-axis.}
 \label{fig:with_attenuation}
\end{figure}

The frequency-domain fit for the scientific data from both magnetic sensors, as elaborated in Section IV, resulted in the findings shown in Fig. \ref{fig:with_attenuation}. The payload's measurements remain unaltered while the FGM signal is reduced in order to match the attenuation of the AMR sensors' magnetic shield. This reduction corresponds to a factor of $\sim$35 for the X-axis, $\sim$455 for the Y-axis, and $\sim$3220 for the Z-axis (for the parameter estimates without temperature subtraction, look at table~\ref{tab:pe}). 

\begin{table}[!t]
\centering
\caption{Parameter estimates ($\mathrm{c_A}$ from Section IV) for the frequency domain fit between the measurements of MELISA-III and the FGM, along with the standard deviation.}
\vspace{-5pt}
\begin{tabular}{ccc}
\hline \hline
\textbf{Axis} & \textbf{\begin{tabular}[c]{@{}c@{}}Parameter estimate\\ without temp. subtraction\end{tabular}} & \textbf{\begin{tabular}[c]{@{}c@{}}Parameter estimate\\ with temp. subtraction\end{tabular}} \\
\hline
X & $0.02864050 \pm 4.88 \times 10^{-6}$ & $0.02864078 \pm 4.87 \times 10^{-6}$ \\
Y & $-0.00220005 \pm 1.71 \times 10^{-6}$ & $-0.00219801 \pm 1.44 \times 10^{-6}$ \\
Z & $-0.00031064 \pm 0.53 \times 10^{-6}$ & $-0.00031133 \pm 0.46 \times 10^{-6}$ \\
\hline \hline
\end{tabular}
\label{tab:pe}
\end{table}

\begin{figure*}[!t]
\centering
 \subfigure[]{\includegraphics[width=1\linewidth]{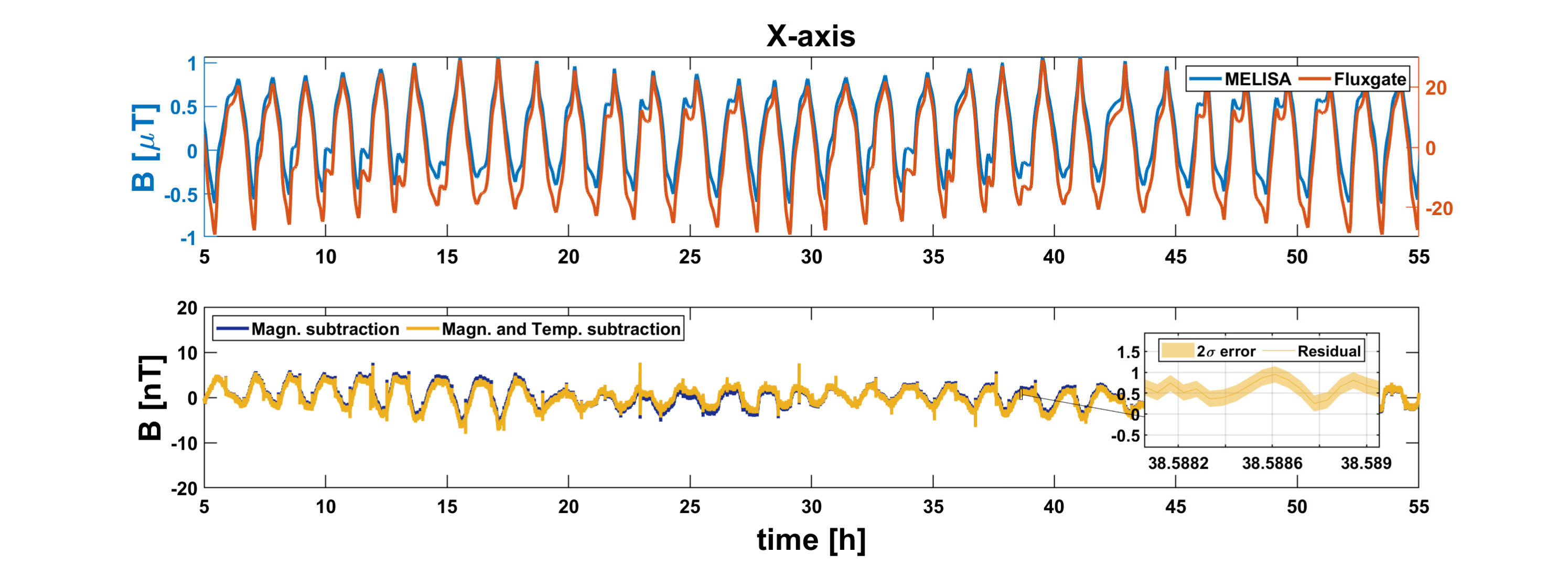}}
 \subfigure[]{\includegraphics[width=1\linewidth]{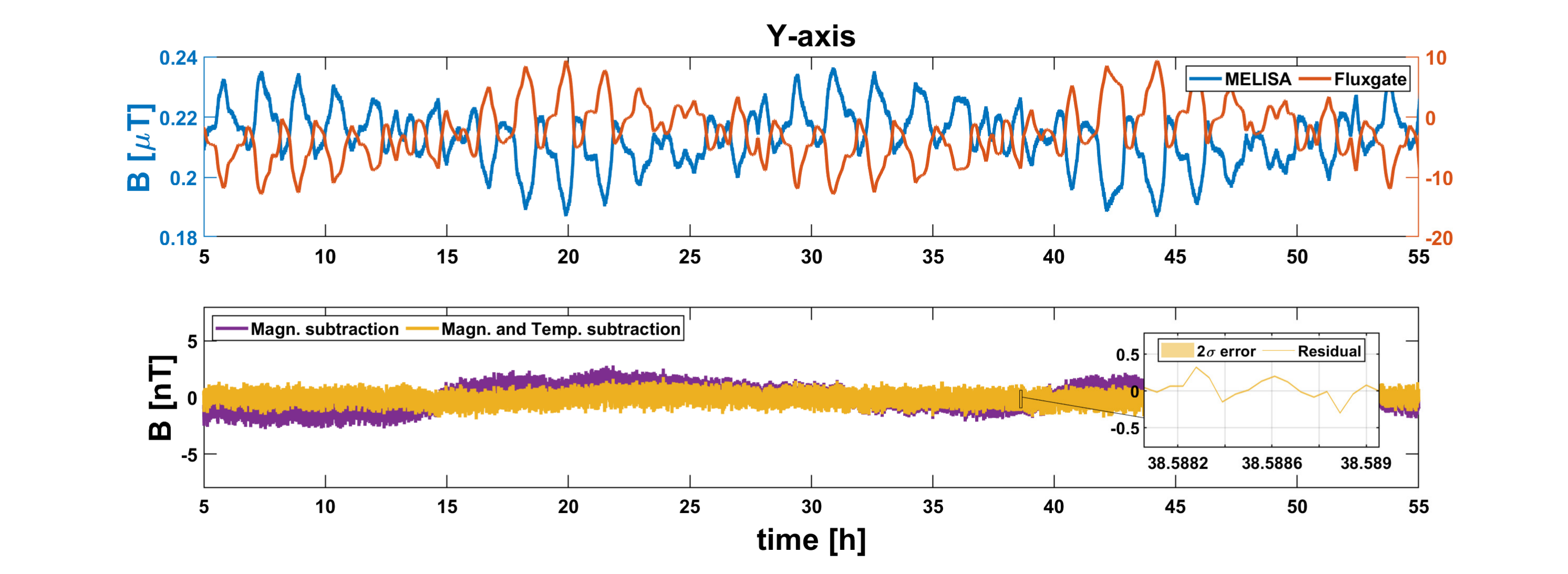}}
 \subfigure[]{\includegraphics[width=1\linewidth]{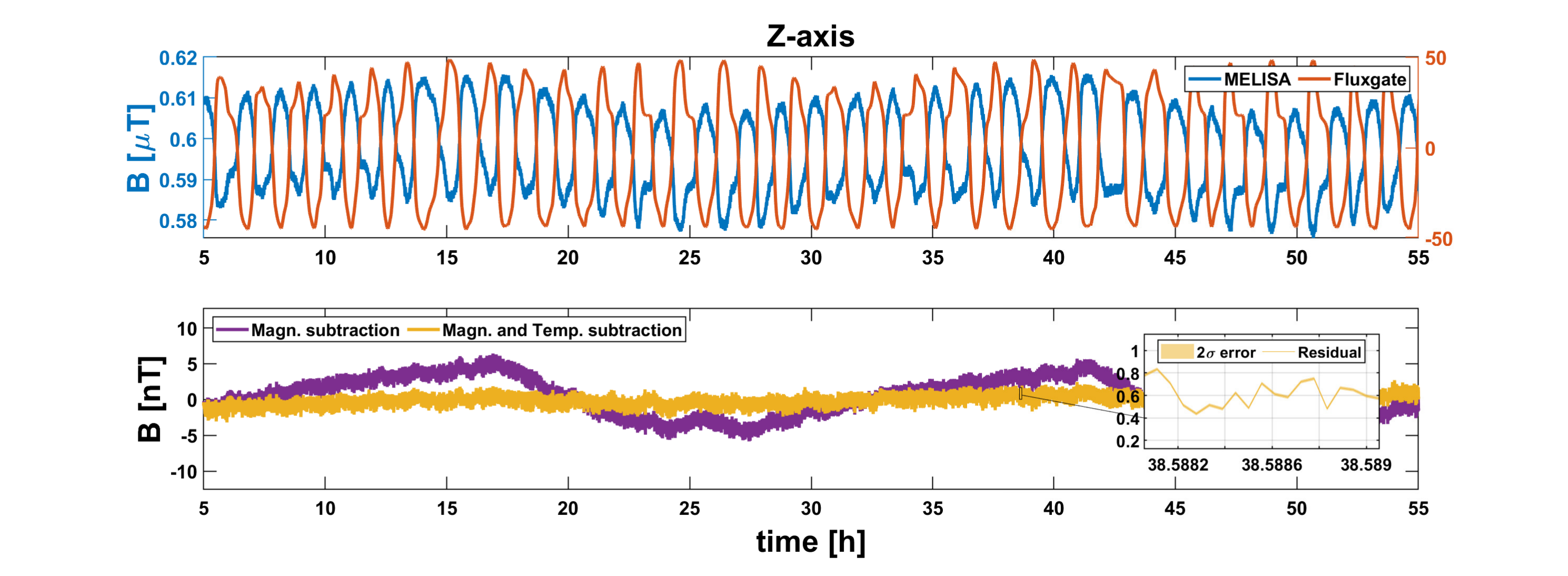}}
 \caption{Magnetic field over time measured by MELISA-III and the FGM, and below that, the residual signals after applying the magnetic and temperature subtraction for the (a) X-axis, (b) Y-axis, and (c) Z-axis. Subtraction was performed at maximum likelihood values for the parameter $\mathrm{c_A}$ recovered by the analysis described in Section ~\ref{sec:analysis}. The $2 \upsigma$ errors are represented by a shaded area surrounding the data.}
 \label{fig:with_without_temp}
\end{figure*}

\begin{figure}[!t]
\centering
 \subfigure[]{\includegraphics[width=1\linewidth]{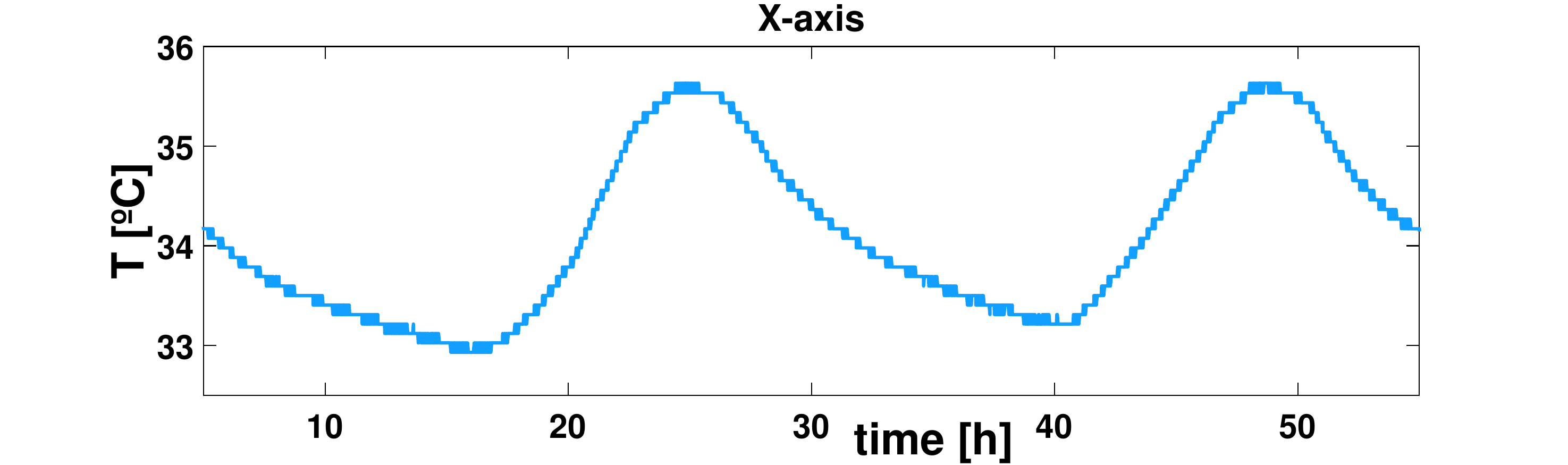}}
 \subfigure[]{\includegraphics[width=1\linewidth]{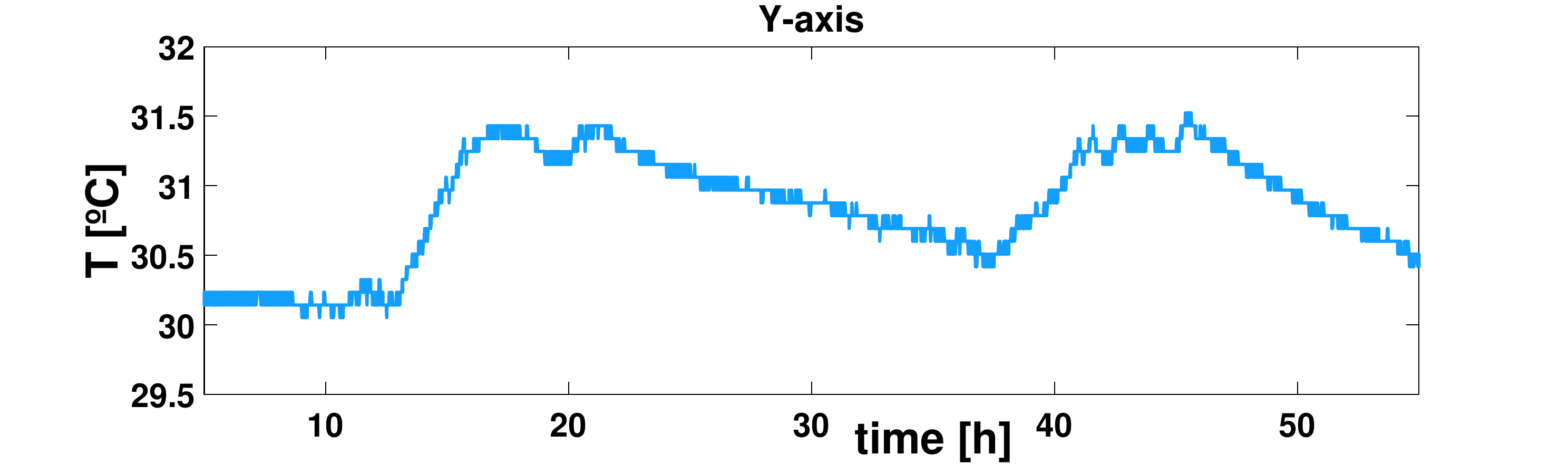}}
 \subfigure[]{\includegraphics[width=1\linewidth]{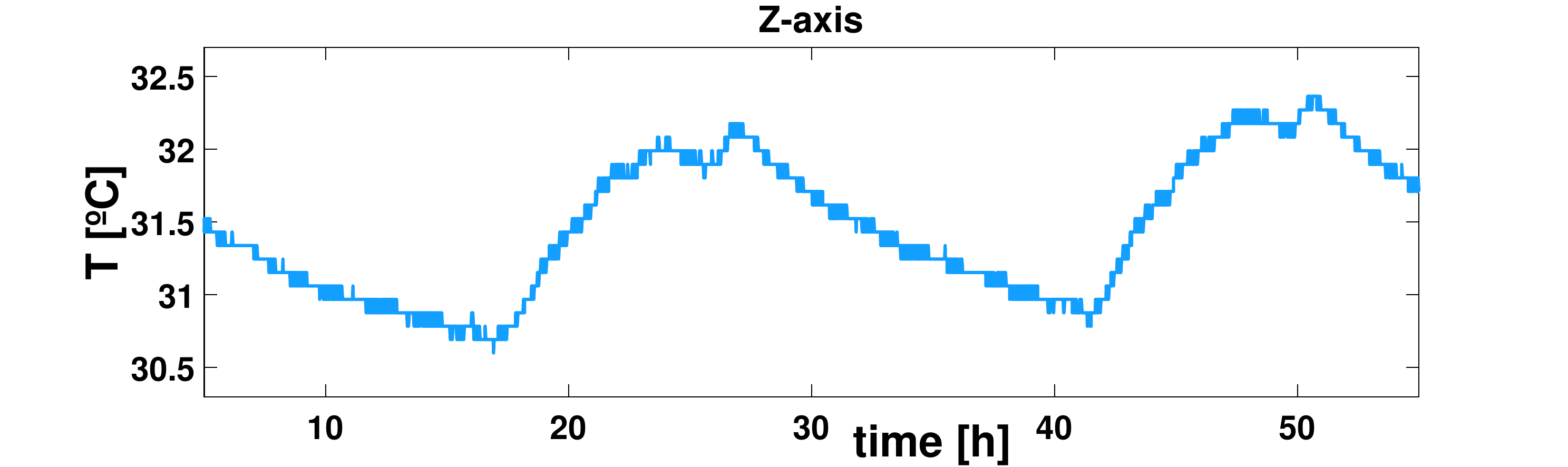}}
 \caption{MELISA-III's NTC thermistor readouts over time during the experiment for the (a) X-axis, (b) Y-axis, and (c) Z-axis.}
 \label{fig:temp}
\end{figure}

The correlated environmental noise is then subtracted from the payload's data, yielding the residual curves displayed in Fig. \ref{fig:with_without_temp}. Nonetheless, two different contributions remain. The first one is the impact of the generated orbit magnetic field, which is still significantly present in the X-axis as cycles every 98 min due to the lower attenuation in this direction. The second contribution is the thermal dependence of the shield, as may be correlated with the temperature measurements in Fig. \ref{fig:temp}. The shield attenuation varies with the ambient thermal changes during the experiment, being the main source of residual noise for the extracted signals in the Y and Z axes with PPMCCs of 0.96428 and -0.96648 regarding the temperature data. For the X-axis, this coefficient is lower, -0.24661, due to the presence of the orbit contribution in the subtracted signal.

\begin{figure}[!t]
\centering
 \subfigure[]{\includegraphics[width=1\linewidth]{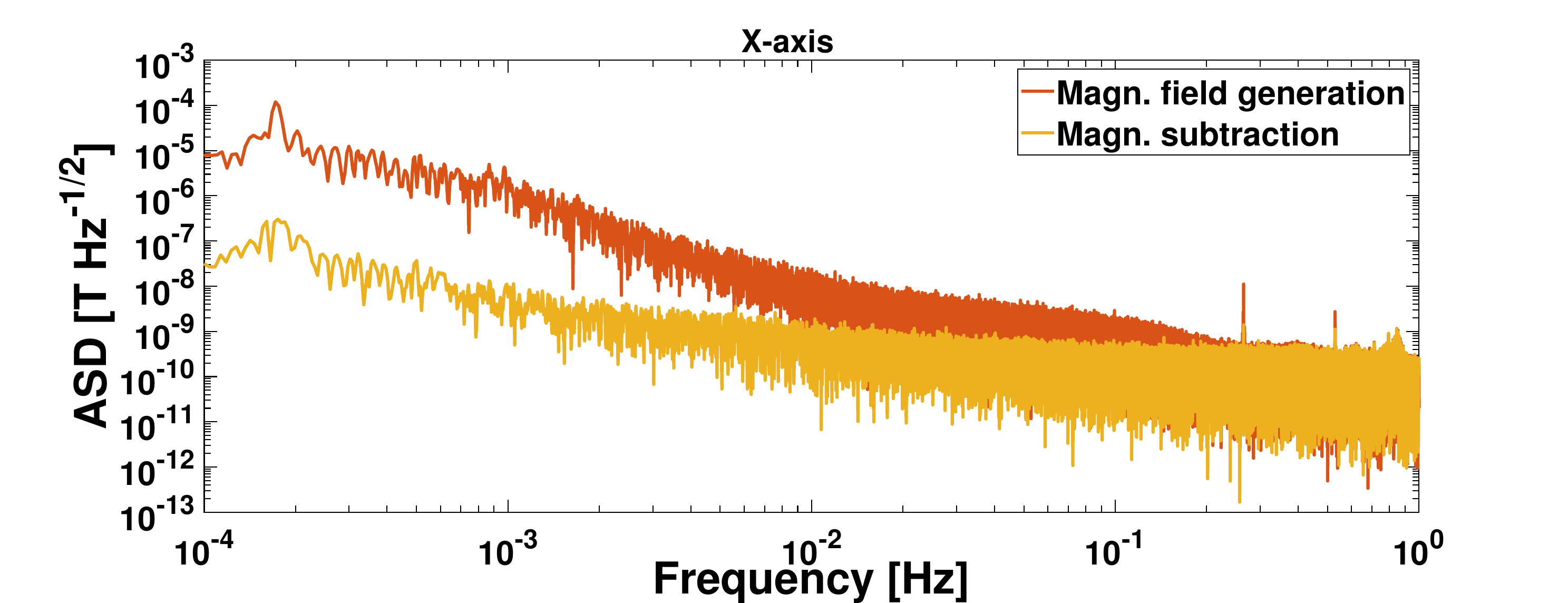}}
 \subfigure[]{\includegraphics[width=1\linewidth]{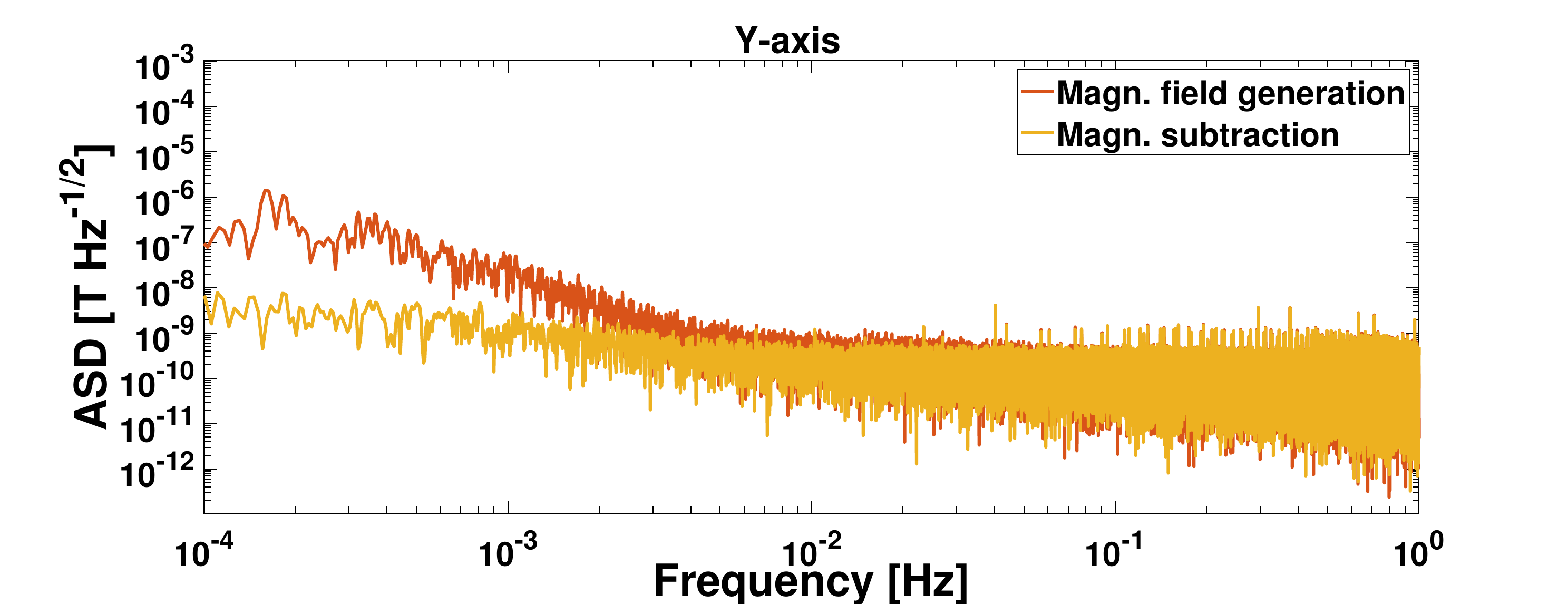}}
 \subfigure[]{\includegraphics[width=1\linewidth]{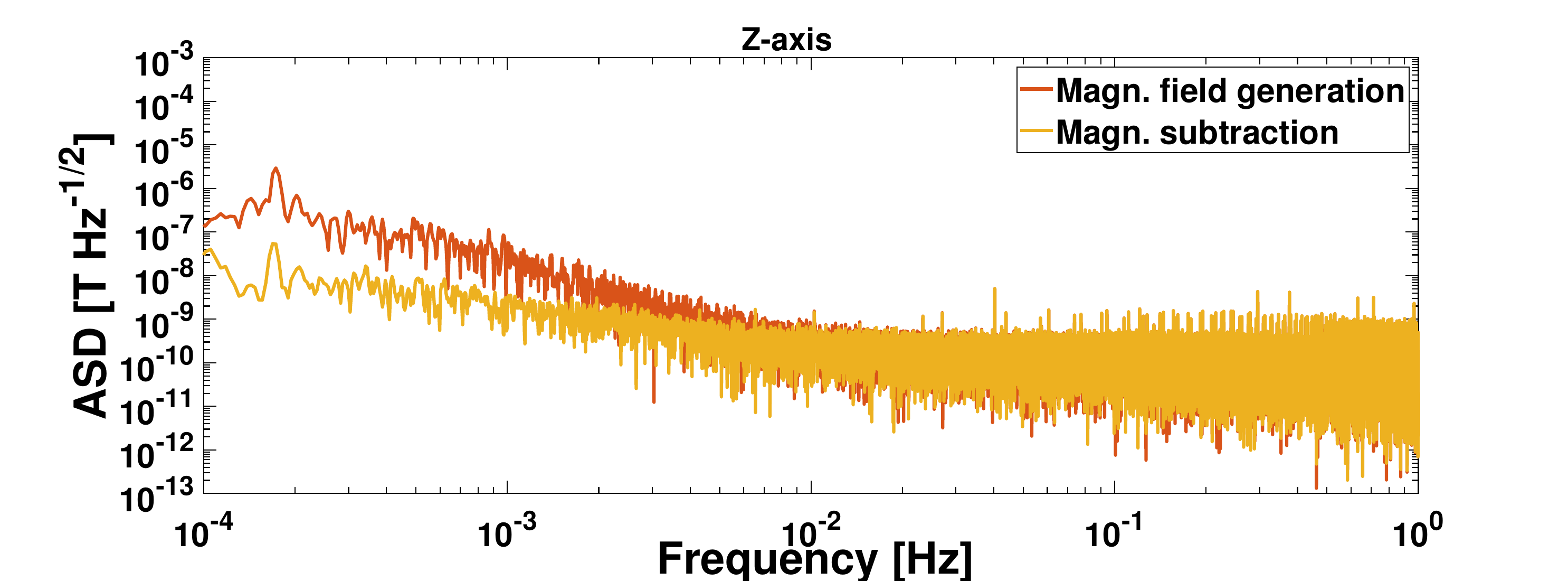}}
 \caption{Comparison between the ASD of MELISA-III's measurements during the orbit simulation and after the magnetic noise subtraction for the (a) X-axis, (b) Y-axis, and (c) Z-axis.}
 \label{fig:asd_extract}
\end{figure}

Thus, a comparison between the ASD of MELISA-III's measurements with and without the correlated magnetic subtraction is displayed in Fig. \ref{fig:asd_extract}. With the elaborated procedure, the low-frequency noise levels have been significantly reduced up to approximately three orders of magnitude around the frequency of the orbital period. The X-axis measurements present the highest residual peak on account of its lower attenuation, which impedes a cleaner subtraction. Still, this fundamental peak, caused by the orbit field generation, has been lowered in this direction from $\mathrm{119\, \upmu T \: Hz^{-1/2}}$ to $\mathrm{302\, nT \: Hz^{-1/2}}$. For the Y and Z axes, the peak has been reduced from $\mathrm{1.3\, \upmu T \: Hz^{-1/2}}$ to $\mathrm{4.3\, nT \: Hz^{-1/2}}$, and from $\mathrm{2.8\, \upmu T \: Hz^{-1/2}}$ to $\mathrm{52.1\, nT \: Hz^{-1/2}}$, respectively.

\begin{figure}[!t]
\centering
 \includegraphics[width=1\linewidth]{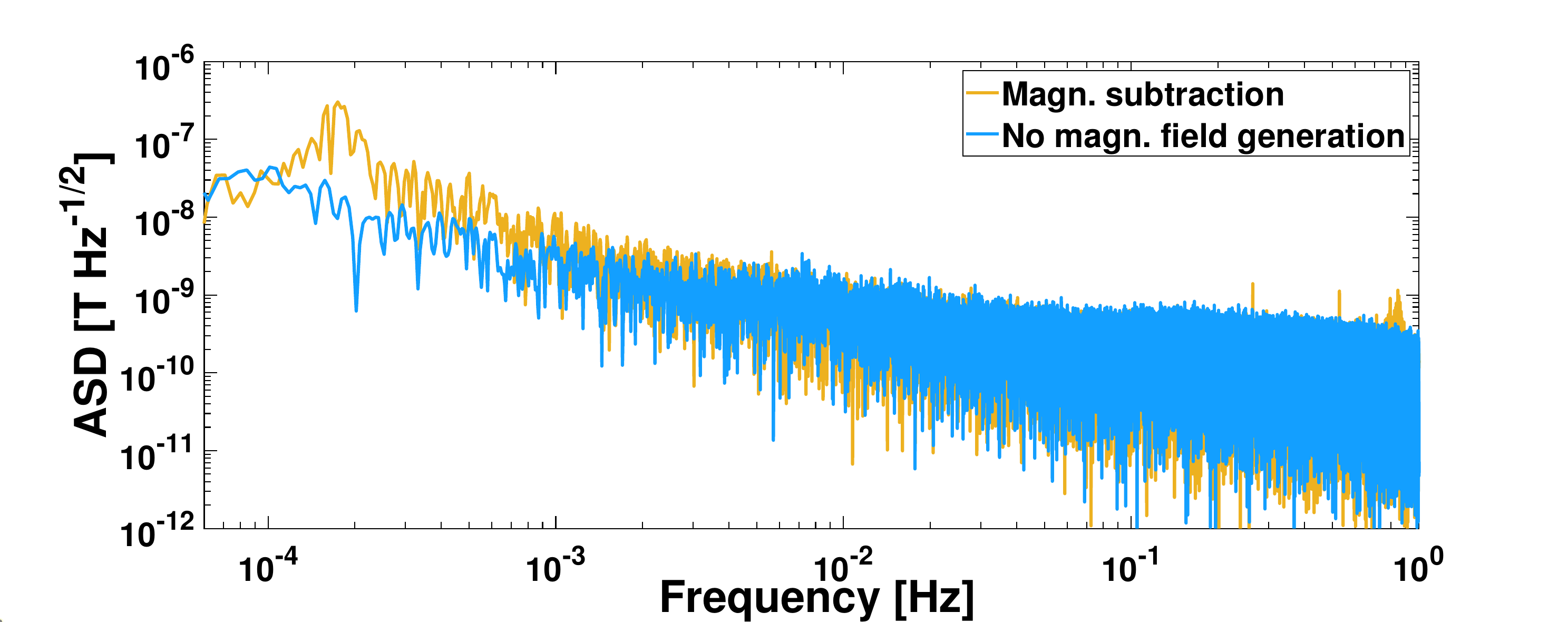}
 \caption{Comparison between the residual ASD after the magnetic subtraction and the shield performance with no magnetic field generation, illustrating a similar drift at low frequencies dominated by the thermal dependence of the material.}
 \label{fig:shield_performance}
\end{figure}

In frequencies different from those influenced by the orbit (around $\mathrm{0.17 \, mHz}$), the current one-layer shield exhibits a similar noise amplitude after subtraction as it did when measuring with no generation of the orbit magnetic environment, as shown in Fig. \ref{fig:shield_performance}. Therefore, the remaining drift in the residual signal below $\mathrm{0.17 \, mHz}$ is not a result of the current subtraction, but a characteristic already present before this experiment as a consequence of the temperature impact on the shield.

\subsubsection{Temperature contribution}

\begin{figure}[!t]
\centering
 \subfigure[]{\includegraphics[width=1\linewidth]{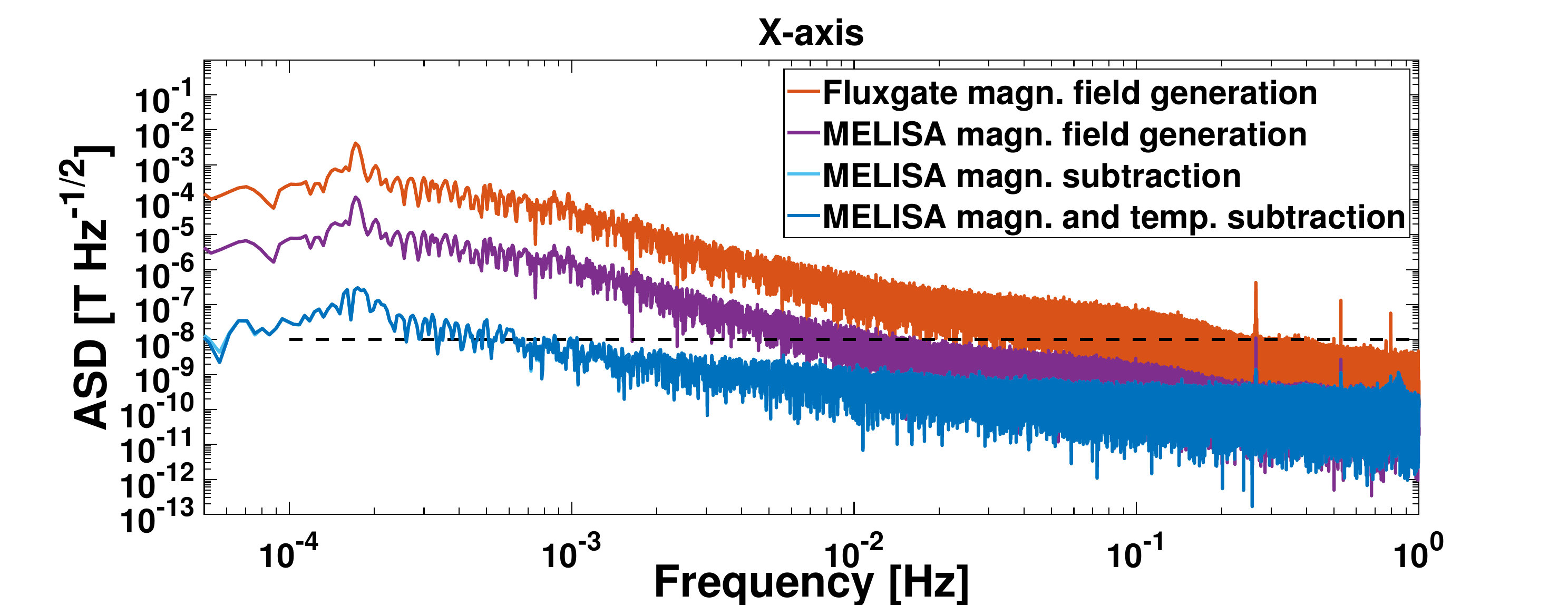}}
 \subfigure[]{\includegraphics[width=1\linewidth]{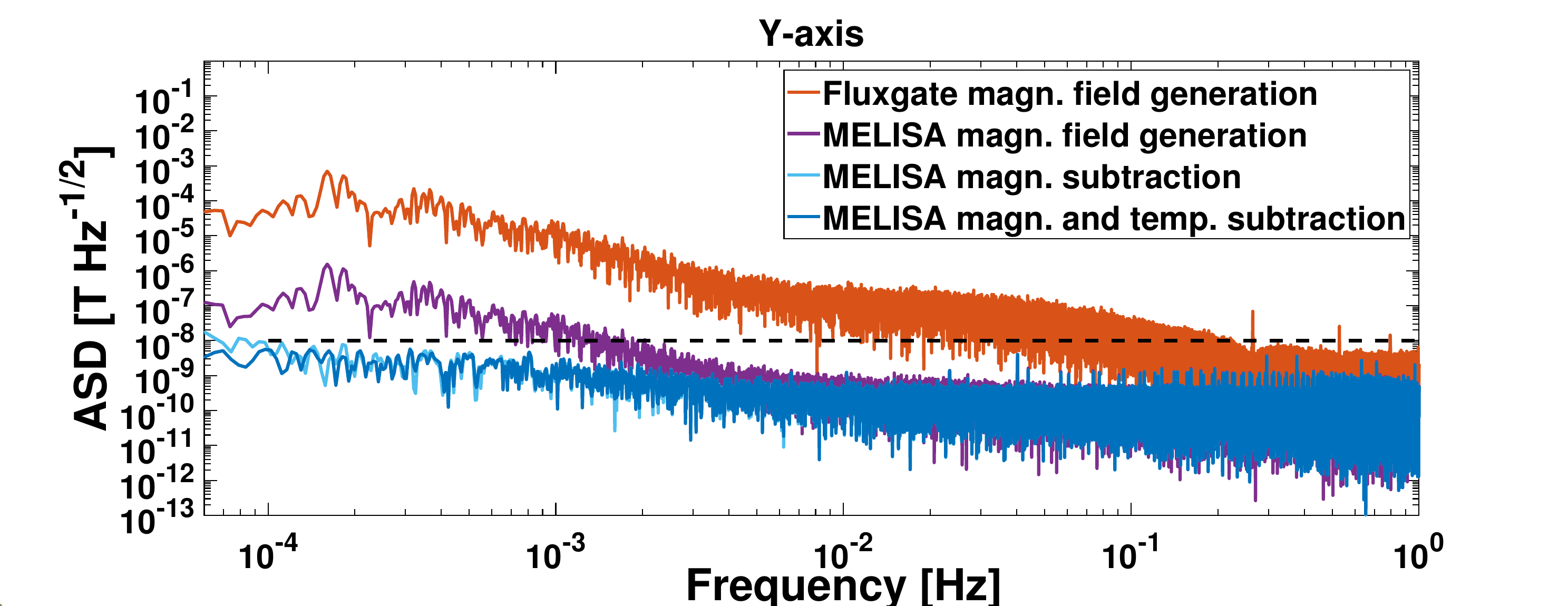}}
 \subfigure[]{\includegraphics[width=1\linewidth]{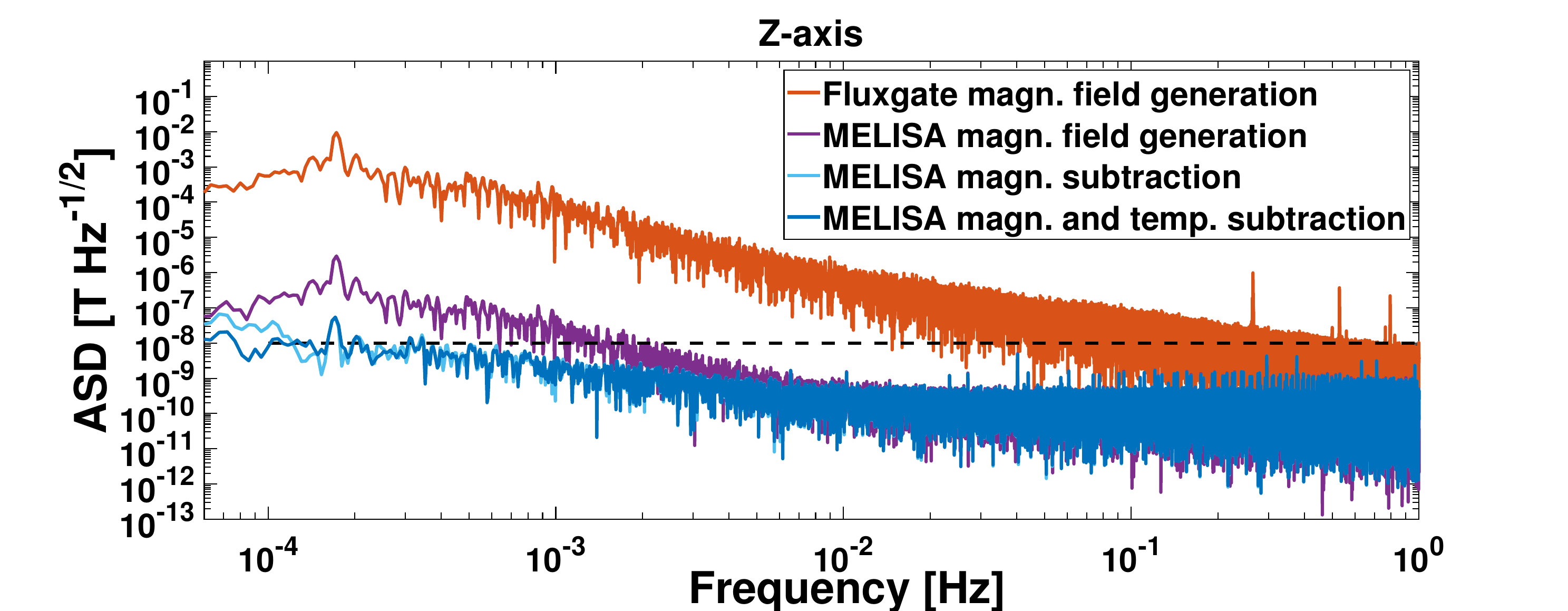}}
 \caption{Comparison between the ASD of MELISA-III's measurements during the generation of the LEO magnetic environment and the different subtraction processes for the (a) X-axis, (b) Y-axis, and (c) Z-axis. Subtraction was performed at maximum likelihood values for the parameter $\mathrm{c_A}$ recovered by the analysis described in Section ~\ref{sec:analysis}.}
 \label{fig:asd_extract_temp}
\end{figure}

The temperature subtraction leads to the residual curves displayed in the time domain in Fig. \ref{fig:with_without_temp}, and in the frequency domain in Fig. \ref{fig:asd_extract_temp}. As a result, magnetic drifts caused by day and night ambient temperature fluctuations are removed from MELISA-III's measurements. In the X-axis, the subtraction is not performed so effectively due to the influence of the generated magnetic field in that direction. The relationship between the variations in the magnetic field and the temperature fluctuations has been considered linear as a first approximation, based on the magnetic oscillations observed in the measurements when the temperature changes.
The elevated value of the PPMCC determined before when the temperature fluctuations are dominant (Y- and Z- axes in Fig. \ref{fig:with_without_temp}) supports this assumption. Nonetheless, further investigation in the thermal characterization of the magnetic shield will allow us to determine a transfer function and hence to enhance the accuracy of the method. In the current work, thermal subtraction is conducted at room temperature to achieve the primary objective of developing a magnetic noise subtraction method.

Overall, the correlation with the temperature data has allowed us to achieve an enhanced noise reduction at the lowest frequencies, especially below the frequency of the orbit (0.17 mHz), in comparison with performing only the magnetic subtraction. Therefore, the combination of both procedures has proved to reduce the noise levels measured by the MELISA-III in the generated SSO magnetic environment by several orders of magnitude.

\subsection{Noise subtraction for a different shielding attenuation}

Considering the magnet onboard the $\Upsigma$yndeo-2 CubeSat, most of its magnetic influence on the payload's measurements is present in the Z-axis. The attenuation in that axis is obtained through the procedure elaborated in Section V, giving a field reduction after integration around 27 dB higher than the calculated for the one-layer shield. For the other axes, the contribution of the magnet is remarkably lower, so the attenuation cannot be obtained precisely due to the remanent magnetic field of the mu-metal shield. Therefore, the previous field reduction obtained for the Z-axis is extrapolated to the others with an uncertainty factor of 20\% \cite{kamens87}, since the mu-metal box covering the payload is expected to exhibit a similar attenuation in the three axes.

\begin{figure}[!t]
\centering
 \includegraphics[width=1\linewidth]{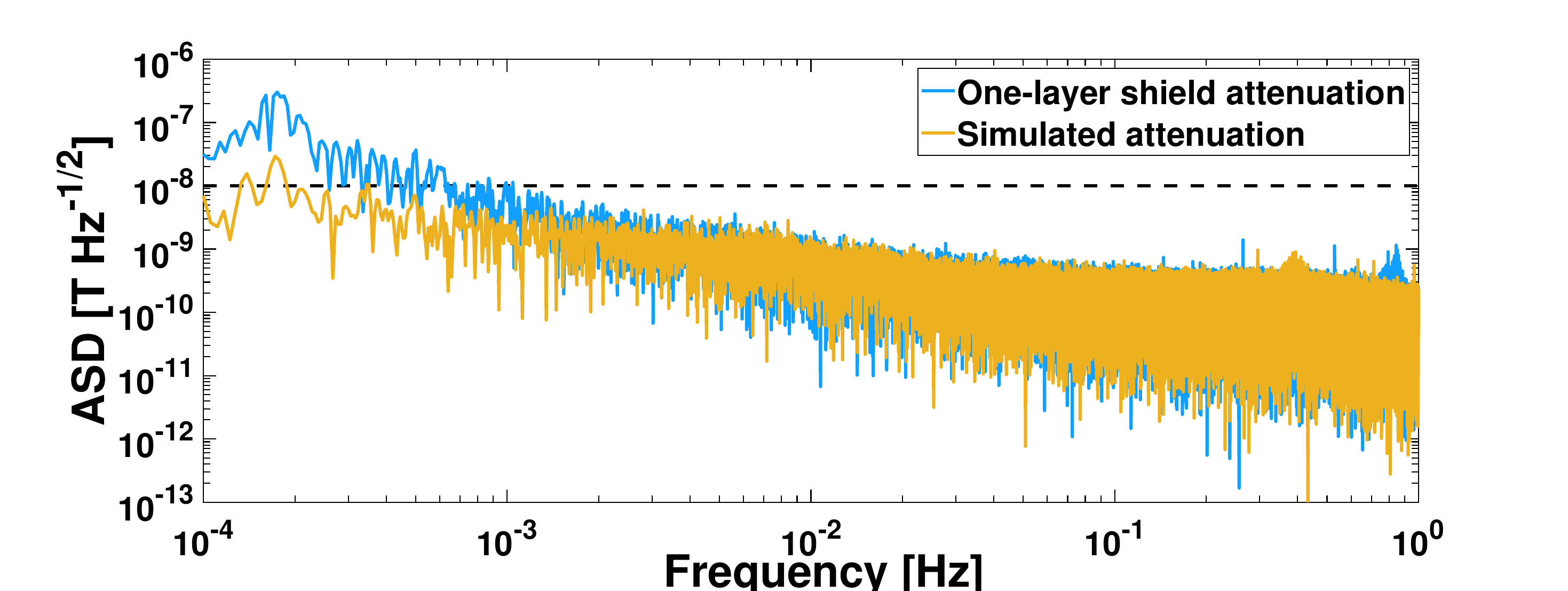}
 \caption{ASD for MELISA-III worst case (X-axis) after the generation of the magnetic field with and without the simulated in-orbit attenuation of the three-layer shield}
 \label{fig:asd_attenuation}
\end{figure}

Finally, applying this attenuation to the magnetic field generated by the Helmholtz coils, the extracted ASD shown in Fig. \ref{fig:asd_attenuation} is obtained for the previous worst case, the X-axis. The subtraction process effectively lowers the noise levels of the orbit fundamental peak down to $\mathrm{29\, nT \: Hz^{-1/2}}$. Despite the 
conservative uncertainty margins that have been considered, the achieved reduction of the residual signal allows to conclude that the present procedure will be effective for the in-orbit setup during operations.

\section{Conclusion}
We have developed a magnetic field simulation setup for the correlation of orbit magnetic noise in a simulated LEO environment. We have subtracted the correlated LEO contributions from the shielded payload's measurements, reducing the noise levels in this environment by up to three orders of magnitude.
We have assessed the performance of this simulator and the noise subtraction procedure down to sub-millihertz frequencies.
Considering also the magnetic attenuation that MELISA-III will experience in orbit, the residual curve gets closer to the mission requirements of magnetic noise levels below $\mathrm{10\, nT \: Hz^{-1/2}}$ in the bandwidth between $\mathrm{0.1 \, mHz}$ and $\mathrm{1 \, Hz}$. In addition, the removal of the thermal dependence of the mu-metal shield has proved to minimize the noise amplitude below the frequency of the orbit (0.17 mHz). The relationship between the magnetic field and the temperature fluctuations has been initially approximated as linear, given the elevated PPMCC values.
Since the range of temperatures experienced in LEO spans from $-20\degree{\rm C}$ to $50\degree{\rm C}$, we plan to replicate the in-orbit thermal environment in future research to enhance our understanding of the relation between temperature and magnetic field variations.

To accomplish a proper noise subtraction, replicating accurately the LEO magnetic conditions with the orbit simulator is crucial. Quantization noise induced by the control of the current provided to the Helmholtz coils, as well as desynchronization between the magnetic sensors, leads to undesired noise contributions which are non-representative of the orbit environment. A smooth field generation resulted in an accurate frequency-domain fit that enables the correlated noise subtraction.

It has been demonstrated during this work that, in addition to the magnetic shield attenuation, further postprocessing is required to reduce the substantial noise caused by the Earth's magnetic field fluctuations undergone in orbit. Consequently, the present procedure will serve as a pivotal tool to mitigate the external contributions and to meticulously evaluate the intrinsic noise of MELISA-III during the in-flight nominal operations that will start in 2024. These outcomes will help to increase the technological maturity of the system for the next generation of space-borne gravitational wave detectors, such as LISA.

\section*{Acknowledgment}
The MELISA-III experiment was embarked on the first mission of the H2020 IOD/IOV program of the EU.

\vskip -2\baselineskip plus -1fil

\begin{IEEEbiography}[{\includegraphics[width=1in,height=1.25in,clip,keepaspectratio]{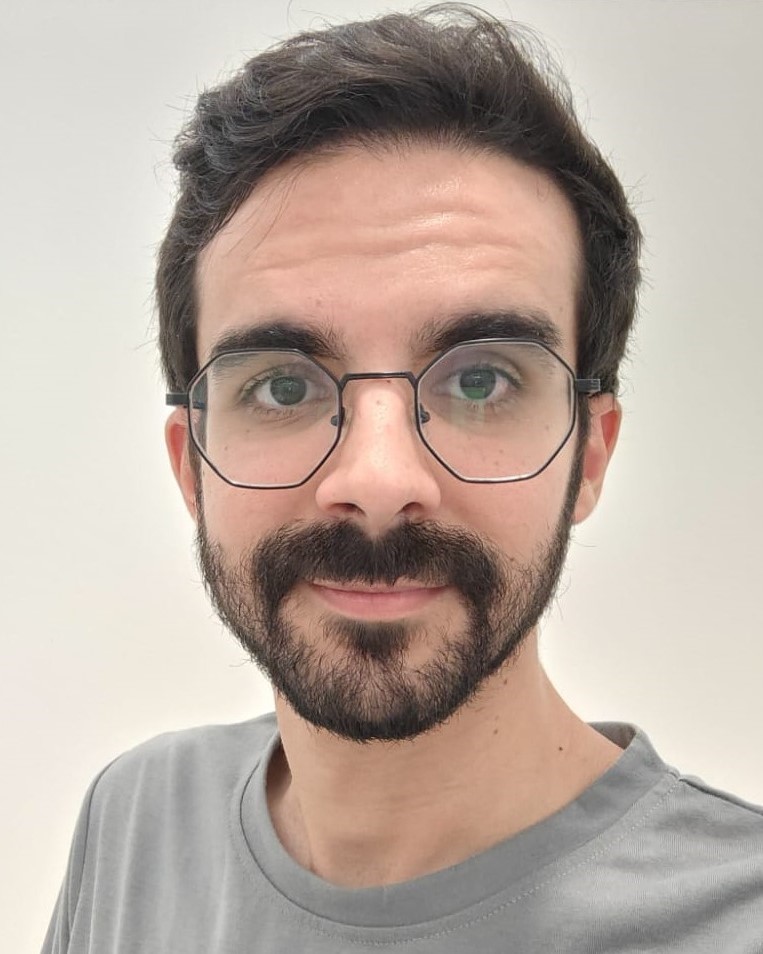}}]{Cristian Maria-Moreno} received the B.S. degree in aerospace engineering and the M.S. in research in systems and computer engineering from the University of Cádiz, Cádiz, Spain, in 2019 and 2021, respectively. He is currently pursuing the Ph.D. degree within the electronic area of that university.

In 2023, he was a Visiting Researcher at the Michigan Physics Department, Ann Arbor, MI, USA, studying atomic magnetometers for space applications. Since 2021, he has been member of the UCAnFly team under the Fly Your Satellite! (FYS!) programme of the European Space Agency (ESA) for the development and launching of a nanosatellite. He is member of the Andalusia LISA
Group of the LISA Mission Consortium. His research focuses on the development of novel magnetometry for space missions.
\end{IEEEbiography}

\vskip -2\baselineskip plus -1fil

\begin{IEEEbiography}[{\includegraphics[width=1in,height=1.25in,clip,keepaspectratio]{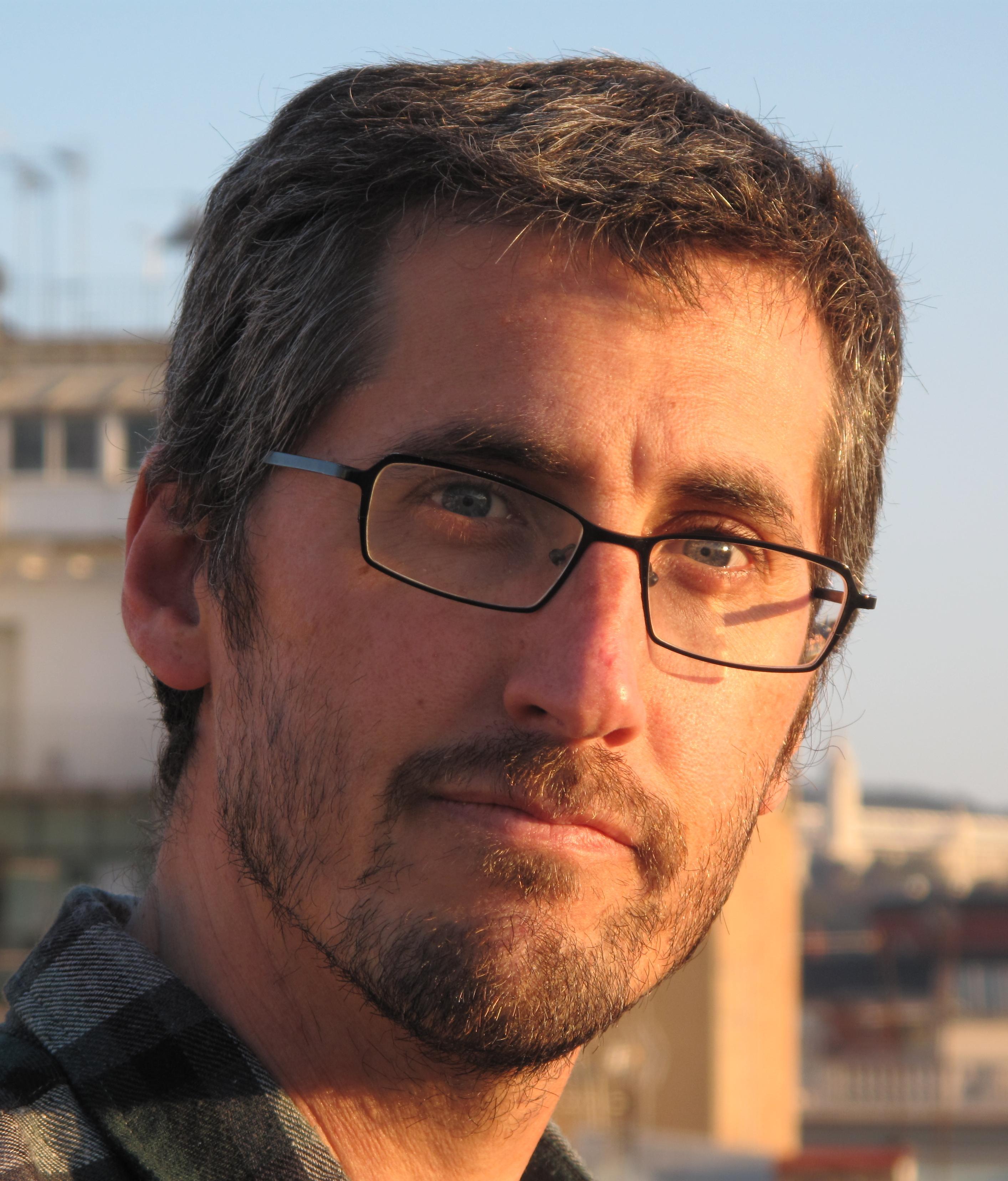}}]{Ignacio Mateos} received the Ph.D. degree in electronics from the Polytechnic University of Catalonia (UPC) in 2015.

During his professional career, he first worked at the Institute of Astrophysics of Andalusia, and then at the Institute of Marine Sciences, focusing on sensors and electronic instrumentation for astronomical and oceanographic research. He was a Research and Development Engineer at the Institute of Space Sciences, and a Visiting Researcher at the Berkeley Physics Department, developing novel techniques in magnetometry for space missions. After a period dedicated to technology transfer from astrophysical instrumentation to medical applications at the Institute of Astrophysics of the Canary Islands, he is now an RyC Researcher at the Universidad de Cádiz.

Dr. Mateos was awarded the Biennial Prize for the best doctoral thesis in astrophysical and astronomical instrumentation by the Spanish Astronomical Society (SEA) in 2017. This work also obtained the Honorable Mention from the Gravitational Wave International Committee (GWIC). He also received the Research Excellence Award in the category of young researcher (area of engineering) from the Universidad de Cádiz in 2019.
\end{IEEEbiography}

\vskip -2\baselineskip plus -1fil

\begin{IEEEbiography}[{\includegraphics[width=1in,height=1.25in,clip,keepaspectratio]{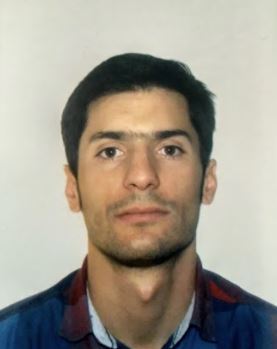}}]{Guillermo Pacheco-Ramos} received his M.Sc degree in Aeronautical Engineering from Escuela Técnica Superior de Ingeniería, Sevilla, Spain, in 2008. With over 6 years of experience as a mechanical engineer in the aeronautics industry, he transitioned to working as an independent consultant specializing in structural analysis and composite materials from 2015 to 2021.

From 2016 to 2021, Guillermo held a part-time assistant professor position at the University of Cádiz, Spain, in the Department of Mechanical Engineering and Design. During this period, he contributed to courses for the B.Sc. Degree in Aerospace Engineering. Since September 2021, he has been assistant professor at the Department of Aerospace Engineering and Fluid Mechanics at the Escuela Técnica Superior de Ingeniería, Sevilla, Spain. 

Pacheco is currently pursuing his Ph.D. thesis on the dynamic and control aspects of multibody systems for aerospace applications. His research interests also include space mission analysis and design, trajectory optimization, and guidance, navigation, and control.
As a member of the Andalusia LISA Group of the LISA Mission Consortium, Guillermo serves as a researcher responsible for the design, analysis, and environmental validation of scientific payloads.
\end{IEEEbiography}

\vskip -2\baselineskip plus -1fil

\begin{IEEEbiography}[{\includegraphics[width=1in,height=1.25in,clip,keepaspectratio]{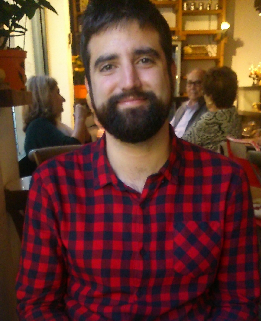}}]
{Francisco Rivas} obtained his PhD in Physics from the Autonomous University of Barcelona in 2019.

Francisco was part of the scientific team of the LISA Pathfinder mission. During this period, he also completed a research stay at the Max Planck Institute for Gravitational Physics (Albert Einstein Institute). Francisco held a postdoctoral position in the Department of Physics at the University of Trento. He is now a lecturer in the Department of Quantitative Methods at Loyola University. Francisco's main research focuses on the study of the technology needed for the observation of gravitational waves in space. 

Rivas-García is a member of the LISA Instrumentation Group, which is part of the Andalusia LISA Group within the LISA mission consortium.
\end{IEEEbiography}

\vskip -2\baselineskip plus -1fil

\begin{IEEEbiography}[{\includegraphics[width=1in,height=1.25in,clip,keepaspectratio]{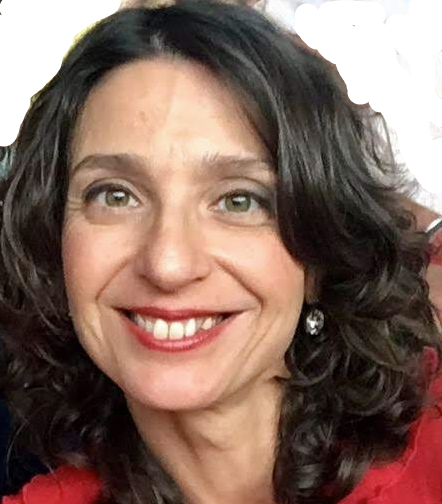}}]{María-Ángeles Cifredo-Chacón} received a B.Sc. degree in Electronic Engineering in 1994 and a B.Sc. degree in Industrial Organization Engineering from the University of Cadiz (Cadiz, Spain) in 1997 and a Ph.D. in Industrial Electronics from the University of Cadiz (Cadiz, Spain) in 2010. 

She has been a Lecturer from 1998 and an Associate Professor from 2016 in the Department of Systems Engineering and Electronics at the University of Cadiz. 
Her research is focused on synthesis of electronic circuits from HDL descriptions in FPGA platform and embedded systems based on microcontrollers.
\end{IEEEbiography}

\vskip -2\baselineskip plus -1fil

\begin{IEEEbiography}[{\includegraphics[width=1in,height=1.25in,clip,keepaspectratio]{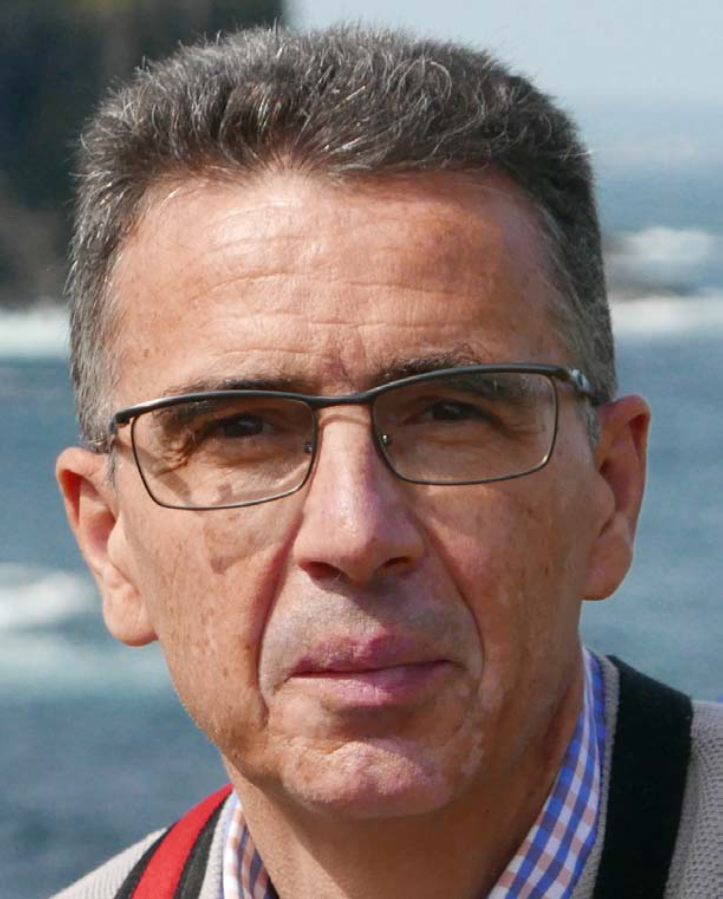}}]{Ángel Quirós-Olozábal} received the B.Sc. degree in electronic engineering from the Universidad de Cádiz, Cádiz, Spain, in 1987, the B.Sc. degree in physics, specialized in electronics, from Universidad Nacional de Educación a Distancia, Madrid, Spain, in 1991, and the Ph.D. degree in industrial electronics from the Universidad de Cádiz in 2002.

He has been an Assistant Professor in electronics from 1987 to 1996 and an Associate Professor since 1996 in electronics with the Department of Systems Engineering and Electronics, Universidad de Cádiz. His research is focused on electronic design, boundary-scan tests, and synthesis of electronic circuits from VHSIC (Very High-Speed Integrated Circuits Program) Hardware Description Language (VHDL).

\end{IEEEbiography}

\vskip -2\baselineskip plus -1fil

\begin{IEEEbiography}[{\includegraphics[width=1in,height=1.25in,clip,keepaspectratio]{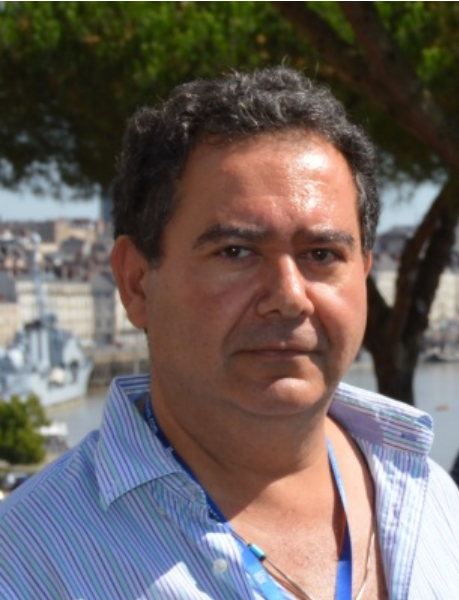}}]{Jose-Maria Guerrero-Rodriguez} received a B.Sc. degree in Electronic Engineering from the University of Cadiz (Spain) in 1987 and a B.Sc. degree in Physics, specialising in Electronics, from UNED University (Madrid, Spain) in 1999. He worked for several electronic sector companies actuating as a test engineer or R\&D engineer. Later, he received a Ph.D. in Industrial Electronics from the University of Cadiz (Cadiz, Spain), in 2009. He joined the Engineering School (University of Cadiz) as a professor in the Electronic Area of the Department of Systems Engineering and Electronics, in 1997. His research is focused on electronic instrumentation and sensor devices and AI techniques application on Intelligent Instrumentation.\end{IEEEbiography}

\vskip -2\baselineskip plus -1fil

\begin{IEEEbiography}[{\includegraphics[width=1in,height=1.25in,clip,keepaspectratio]{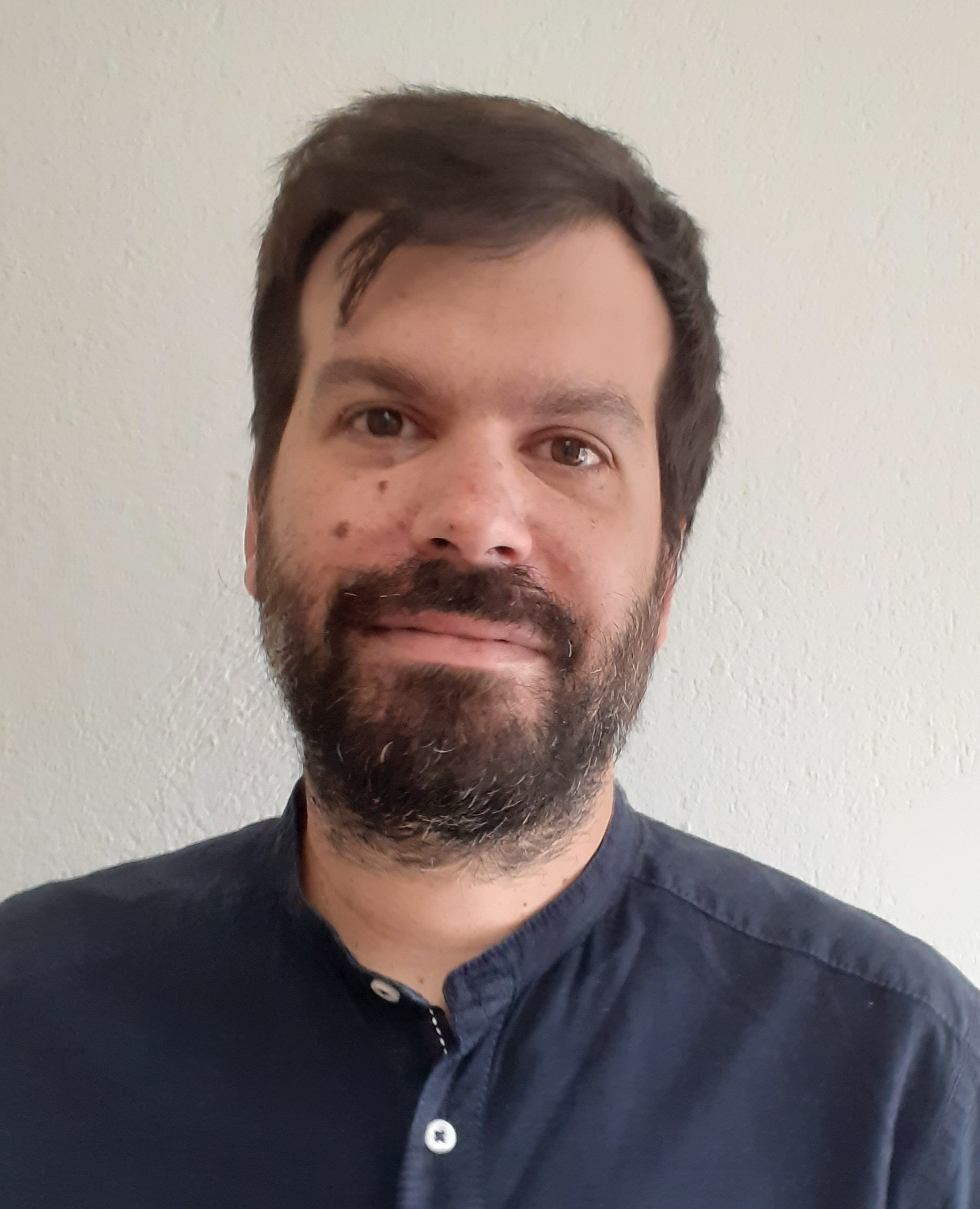}}]{Nikolaos Karnesis} graduated from the School of Applied Mathematical and Physical Sciences of the National Technical University of Athens. Obtained PhD from the Universitat Autònoma de Barcelona and the Institut d'Estudis Espacials de Catalunya.

He worked as a postdoctoral researcher at the Max Planck Institute for Gravitational Physics (Albert Einstein Institute in Hannover), and at the laboratoire Astroparticule et Cosmologie (APC Paris). Currently based at the Physics Department of the Aristotle University of Thessaloniki in Thessaloniki, Greece.

Dr. Karnesis is member of the LISA, the LISA Science Team, and the LVK collaboration, the Hellenic Society of Relativity Gravitation and Cosmology, and the Astronomy \& Fundamental Physics Panel of the European Space Sciences Committee.
\end{IEEEbiography}

\end{document}